\documentclass[journal,twoside]{IEEEtran}
\usepackage{color}
\usepackage{subfigure}
\usepackage{graphicx}
\usepackage{booktabs}
\usepackage{longtable}
\usepackage{amssymb}
\usepackage{amsmath}
\usepackage{cite}
\usepackage{multirow}
\usepackage{url}
\usepackage{tabularray}
\usepackage{pgfplots}
\pgfplotsset{compat=1.5}
\usepgfplotslibrary{groupplots}
\usepackage{algorithm}
\usepackage{algpseudocode}
\usepackage{amsmath}
\renewcommand{\algorithmicrequire}{\textbf{input:}}
\renewcommand{\algorithmicensure}{\textbf{output:}}

\usepackage{colortbl}
\usepackage{algorithmicx}
\usepackage{nicematrix}
\usepackage{threeparttable}    
\usepackage{listings}
\usepackage{fontawesome}

\usepackage[switch, pagewise]{lineno}

\usepackage[
    colorlinks=true, 
    linkcolor=red,   
    urlcolor=blue,    
    citecolor=blue    
]{hyperref}

\begin{document}

\title{Partition Map-Based Fast Block Partitioning for VVC Inter Coding}

\author{
Xinmin Feng,
Zhuoyuan Li, 
Li Li, \IEEEmembership{Member, IEEE},
Dong Liu, \IEEEmembership{Senior Member, IEEE},
and Feng Wu, \IEEEmembership{Fellow, IEEE}
\thanks{
Date of current version \today. 
This work has been submitted to the IEEE for possible publication.  
Copyright may be transferred without notice, after which this version may no longer be accessible.

The authors are with the MOE Key Laboratory of Brain-Inspired Intelligent Perception and Cognition, University of Science and Technology of China, Hefei 230027, China.
(e-mail: xmfeng2000@mail.ustc.edu.cn; zhuoyuanli@mail.ustc.edu.cn;  lil1@ustc.edu.cn; dongeliu@ustc.edu.cn; fengwu@ustc.edu.cn).
}
}

\markboth{Accepted by IEEE Transactions on MultiMedia}%
{Feng \MakeLowercase{\textit{et al.}}: Partition Map-Based Fast Block Partitioning for VVC Inter Coding}

\maketitle

\begin{abstract}
Among the new techniques of Versatile Video Coding (VVC), the quadtree with nested multi-type tree (QT+MTT) block structure yields significant coding gains by providing more flexible block partitioning patterns. However, the recursive partition search in the VVC encoder increases the encoder complexity substantially. To address this issue, we propose a partition map-based algorithm to pursue fast block partitioning in inter coding. Based on our previous work on partition map-based methods for intra coding, we analyze the characteristics of VVC inter coding, and thus improve the partition map by incorporating an MTT mask for early termination. Next, we develop a neural network that uses both spatial and temporal features to predict the partition map. It consists of several special designs including stacked top-down and bottom-up processing, quantization parameter modulation layers, and partitioning-adaptive warping. Furthermore, we present a dual-threshold decision scheme to achieve a fine-grained trade-off between complexity reduction and rate-distortion (RD) performance loss. The experimental results demonstrate that the proposed method achieves an average 51.30\% encoding time saving with a 2.12\% Bjøntegaard Delta Bit Rate (BDBR) under the random access configuration.
The source code could be available at \url{https://github.com/ustcivclab/IPM}.
\end{abstract}


\section{Introduction} \label{introduction}
\IEEEPARstart{W}{ith} the advancement of display  technology, the demand for compressing high-resolution videos has increased significantly. To address this, the Joint Video Experts Team (JVET) developed Versatile Video Coding (VVC) \cite{bross2021overview}.
Among the various new technical aspects introduced in VVC, the quadtree plus multi-type tree (QT+MTT) block partitioning structure has been identified as one of the most significant changes compared with the High Efficiency Video Coding (HEVC) standard \cite{sullivan2012overview, huang2021blockstructure}. Specifically, the JVET test software (JEM) adopts the quadtree plus binary tree (QTBT) partitioning structure to better adapt to diverse texture characteristics \cite{JEM2025}. The VVC test model (VTM) further enhances QTBT by introducing two ternary partitioning modes, enabling more flexible coding unit (CU) partition patterns.
However, this significantly increases encoder complexity, as the optimal CU partition is determined through a brute-force rate-distortion optimization (RDO) search.
Specifically, under the random access (RA) configuration, VTM incurs a 617\% increase in encoding time compared with the HEVC model (HM) under default settings \cite{bossen2021complexity}.
Therefore, it is necessary to accelerate the VVC encoder while preserving a desirable coding efficiency.

Considerable effort has been devoted to reducing the encoding complexity of VVC intra and inter coding, including handcrafted feature-based methods  \cite{dong2022intra}, \cite{amestoy2019random}, \cite{yang2019low},  \cite{lei2019intra}, \cite{wu2021intra}, \cite{cui2020gradient}, \cite{chen2019intra}, \cite{Saldanha2021}, \cite{fastvtm}, \cite{tang'18}, \cite{huang2023precise}, \cite{amestoy2019tunable} and neural network-based approaches \cite{jin2017cnn}, \cite{galpin2019cnn}, \cite{lty2021intra}, \cite{wu2022hg}, \cite{Tissier2023Intra}, \cite{fal2023}, \cite{pan'21}, \cite{Tissier'22}, \cite{peng2023classification}, \cite{lin2024efficient}, \cite{park2020fast}.
By utilizing effective partition representations, neural network-based methods achieve higher acceleration ratios with lower compression loss in VVC intra coding compared to traditional approaches. However, in contrast to the progress made in intra coding, reducing the complexity of inter coding remains a more challenging task. Specifically, fast block partitioning algorithms for VVC inter coding can be improved in three key aspects: (1) developing an effective representation of the QT+MTT partition structure that reflects the characteristics of inter coding, (2) designing a neural network capable of modeling the complex relationship between pixel-level features and block partitioning in inter coding, and (3) implementing a flexible post-processing algorithm to achieve a fine-grained balance between complexity reduction and rate-distortion (RD) performance.

In this paper, we propose a partition map-based algorithm for accelerating block partitioning in VVC inter coding. Building on our previous work in VVC intra coding \cite{fal2023}, we extend the method to inter coding through three key components: representation, neural network architecture, and post-processing, corresponding to Sections~\ref{sec:improved partition map}, \ref{sec:Architecture}, and \ref{sec:post-processing}, respectively.
For the \textbf{representation}, we analyze the characteristics of VVC inter coding and enhance the partition map by introducing an MTT mask to enable early termination of unnecessary MTT splits.
Regarding the \textbf{neural network architecture}, we propose a coarse-to-fine prediction framework in which higher-resolution inputs correspond to the partitioning of smaller blocks. Inspired by the partition search process, we introduce several novel modules, including a slice quantization parameter (QP) modulation layer that adapts to varying compression levels across frames, and a partitioning-adaptive warping module that emulates block-based motion estimation. Unlike prior designs, we elevate QT depth map prediction from the coding tree unit (CTU) level to the frame level, thereby leveraging global motion cues derived from optical flow estimation~\cite{Ranjan_2017_CVPR_spynet}. However, ensuring the compliance of the prediction results with partitioning rules remains a key challenge for partition map prediction. To address this, we introduce a progressive refinement strategy that leverages repeated top-down and bottom-up processing through stacked hourglass blocks with intermediate supervision. This strategy encourages neural networks to develop a coherent understanding of the partition structure, thereby improving the local consistency of the predicted results.
For \textbf{post-processing}, we introduce a dual-threshold decision scheme that achieves a fine-grained trade-off between encoding complexity and RD performance.
Experimental results on the JVET common test sequences \cite{Bossen2020sequence} demonstrate that the proposed method reduces inter coding time by 51.30\%, with only a 2.12\% increase in Bjøntegaard Delta Bit Rate (BDBR), outperforming existing approaches.
In summary, the main contributions of this paper are as follows:

\begin{itemize} \item We enhance the partition map to efficiently represent the QT+MTT partition structure in VVC inter coding by introducing the MTT mask.

\item Inspired by the partition search process in inter coding, we design a novel neural network that predicts the partition map in a coarse-to-fine manner. The network integrates several new modules, including repeated top-down and bottom-up processing, a partitioning-adaptive warping module, and a QP modulation layer.

\item We propose a dual-threshold decision scheme that enables fine-grained control over the trade-off between encoding complexity and RD performance. 

\item Experimental results show that our approach achieves an advanced trade-off between encoding time reduction and BDBR performance under the RA configuration.

\end{itemize}

\section{Related Work} \label{sec:related work}

In this section, we briefly review the previous fast block partitioning algorithms used in HEVC and VVC standards.

\subsection{Fast Block Partitioning Algorithms for HEVC Standard}

Before VVC, HEVC has been widely adopted in practical scenarios. Thus, numerous techniques are employed to reduce the complexity of the QT partition search of HEVC in both intra coding and inter coding. 
With respect to the intra coding of HEVC, the relevant methods can be classified into two categories: heuristic methods 
\cite{Cho2013intra}, \cite{Kim2016}, \cite{CHIANG201913}, \cite{fang2016intra}, \cite{Zhang2019intrafast} 
and neural network-based methods \cite{liu2016intra}, \cite{tissier2020intra}, \cite{xu2018intra}. For heuristic methods, researchers 
\cite{Cho2013intra}, \cite{Kim2016}, \cite{CHIANG201913}, \cite{fang2016intra}, \cite{Zhang2019intrafast} 
utilized the intermediate characteristics of the current CU and the spatial correlations with neighboring CUs to  early terminate the unnecessary partition search. 
After this, researchers focused on the effective representation of partition search and predicted it with the deep neural networks. For instance, 
Liu \textit{et al.} \cite{liu2016intra} proposed a VLSI-friendly fast algorithm using convolutional neural networks. Xu \textit{et al.} \cite{xu2018intra} proposed an early terminated hierarchical CNN for learning to predict the hierarchical CU partition map. Tissier \textit{et al.} \cite{tissier2020intra} proposed a probability vector for each 64 $\times$ 64 CU to speed up block partitioning.  Feng \textit{et al.} \cite{fal2019hm} used the depth map to represent the block partition of a CTU and designed a CNN to predict the depth map.

In HEVC inter coding, researchers explored temporal features to early-terminate the redundant partition searches.  
Correa \textit{et al.} \cite{Correa2015inter} extracted intermediate data and built three sets of decision trees. Zhang \textit{et al.} \cite{zhang2015inter} designed a three-output joint classifier considering nine features relevant to CU depth decision. Zhu \textit{et al.} \cite{zhu2017inter} proposed a binary and multi-class SVM algorithm to predict the CU partition with an off-on-line machine learning mechanism. 
Xu \textit{et al.} \cite{xu2018intra} proposed  an early-terminated hierarchical long- and short-term memory network to learn the temporal correlation of the CU partition.

\begin{figure*}[t]
	\centering 
    \includegraphics[trim=0cm 0cm 0cm 0cm, clip, width=0.95\textwidth]{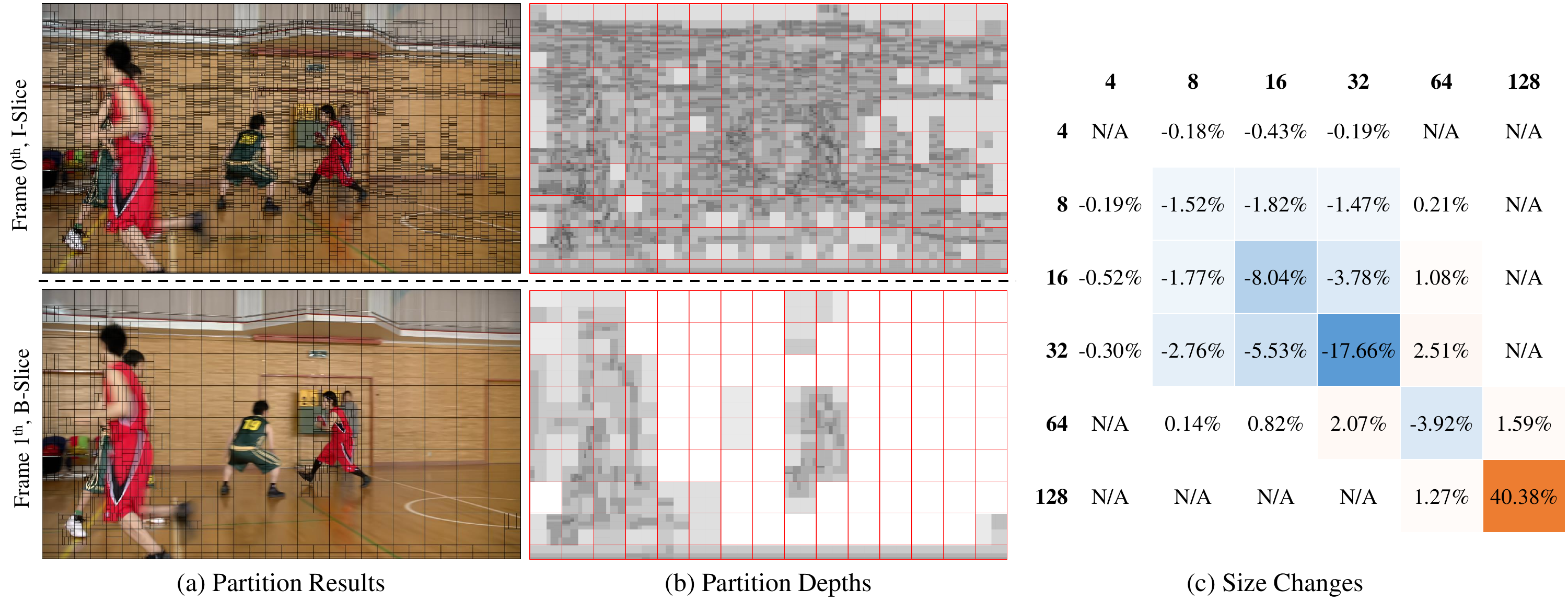}
	\caption{\textbf{(a) Partition results of the I-Slice and B-Slice.}
\textbf{ (b) Partition depths of the two slices.}The red solid lines are the boundaries of CTUs. The darker colors signify deeper partition depths.
\textbf{(c) Changes in the Proportion of Pixels Associated with Different CU Sizes in VVC Inter-Slice Compared to Intra-Slice.}
 Each square indicates a certain size of CUs, with their height and width marked on the left and top.
 “N/A” indicates not available.
} 
	\label{fig:improved partition map}
\end{figure*}

\subsection{Fast Block Partitioning Algorithms for VVC Standard}

With the introduction of the QT+MTT structure, VVC provides more flexible partitioning patterns compared to HEVC, but it also significantly increases encoding time. To address this, various approaches have been developed to speed up the VVC encoding process for both intra and inter coding.

For intra coding of VVC, earlier work focused on  handcrafted feature engineering for fast block partitioning \cite{dong2022intra}, \cite{amestoy2019random}, \cite{yang2019low},  \cite{lei2019intra}, \cite{wu2021intra},   \cite{cui2020gradient}, \cite{chen2019intra},  \cite{Saldanha2021}, \cite{yang2020intra}. Notably, Saldanha \textit{et al.} \cite{Saldanha2021} proposed a configurable light gradient boosting machine that uses effective texture, coding, and context features to predict the best partition type. After this, researchers attempted to model the QT+MTT structure in effective representations and predict them using powerful neural networks.
For instance, 
Galpin \textit{et al.} \cite{galpin2019cnn} modeled the CU boundaries as a single vector, and designed a convolution neural network to predict the vector in a bottom-up manner.
Li \textit{et al.} \cite{lty2021intra} categorized all possible CU sizes into six stages and designed an MSE-CNN to determine the CU partition. 
Wu \textit{et al.} \cite{wu2022hg} devised a hierarchy grid map to represent the QT+MTT structure and proposed HG-FCN to predict it in a stage-wise top-down manner.
\textcolor{black}{Park \textit{et al.} \cite{park2020fast} proposed a lightweight neural network that decides whether to terminate the nested ternary tree following a quadtree, based on both explicit and derived VVC features.}
Tissier \textit{et al.} \cite{Tissier2023Intra} proposed a decision tree model to predict the probabilities at each block of the entire CTU. Feng \textit{et al.} \cite{fal2023} formulated the QT+MTT structure as a partition map and designed a Down-Up-CNN to predict it. 

Regarding inter coding of VVC, fast block partitioning algorithms can also be categorized into heuristic \cite{fastvtm}, \cite{tang'18}, \cite{huang2023precise} and learning-based \cite{amestoy2019tunable}, \cite{pan'21}, \cite{Tissier'22}, \cite{peng2023classification}, \cite{lin2024efficient} methods. For heuristic methods, Wieckowski \textit{et al.} \cite{fastvtm} introduced some practical tools, such as split cost-based early-termination, content-based gradient speed-up, and residual-based TT split prohibition etc. 
Besides, Huang \textit{et al.} \cite{huang2023precise} implemented a scheme that allowed for precisely accelerating the encoding process within one pass. 
For learning-based methods, most research has focused on exploiting temporal motion features with deep learning tools. \textcolor{black}{Amestoy \textit{et al.} \cite{amestoy2019tunable} introduced the first complexity reduction for the VTM reference software in the inter coding configuration using random forest classifiers.} Pan \textit{et al.} \cite{pan'21} proposed a CNN scheme by fusing features extracted from the luma component, residue, and motion field. However, their approach could only handle a subset of CUs, which limited the acceleration ratio.  Then, Tissier \textit{et al.} \cite{Tissier'22} utilized MobileNetV2 \cite{MobileNetV2} as the backbone, which took the current CTU and reference CTUs as input, and predicted a vector that represents all possible  CTU partition structures. 
Furthermore, Peng \textit{et al.} \cite{peng2023classification} modeled the structure as a partition homogeneity map (PHM). 
However, the PHM lacked a tunable complexity reduction mechanism for QT+MTT partitioning decisions, which limited the application scenarios.
In conclusion, by utilizing effective partition representations, neural network-based methods achieved a higher acceleration ratio with lower compression loss compared to other approaches.

\section{Representation of the Partition Structure} \label{sec:improved partition map}

In this section, we explore the differences between the partition results of VVC intra and inter coding, and then enhance the partition map by integrating early-termination mechanisms based on statistical analysis.

\subsection{The Observations of the Partition Results}

To investigate the differences in block partitioning between VVC intra and inter coding,
we compress the JVET test sequences using VTM-10.0 at various quantization parameters, and analyze the differences in CU size distribution between intra and inter coding. 
Taking \textit{BasketballDrive\_1920$\times$1080} as an example, Fig. \ref{fig:improved partition map}(a) shows the partition results for the first two frames, i.e., an I-Slice and a B-Slice. 
While the visual content is similar, CUs in B-Slices tend to be split more coarsely than those in I-Slices, especially since early-terminated CTUs are common in B-Slices but rare in I-Slices.
Furthermore, we analyze the proportions of pixels associated with different CU sizes between VVC intra and inter coding, and present the changes in Fig. \ref{fig:improved partition map}(b). Each square represents a certain size of CUs, with their height and width marked on the left and top. 
Negative values in blue indicate a decrease in the proportion of pixels belonging to CU sizes in the inter-slice compared to the intra-slice, whereas positive values in orange signify an increase in proportion. 
The statistics show that the proportion of pixels associated with 128$\times$128 CTUs increases significantly in inter coding compared to that in intra coding. 

\begin{figure}[t]
	\centering 
    \includegraphics[trim=0cm 0cm 0cm 0cm, clip, width=0.45\textwidth]{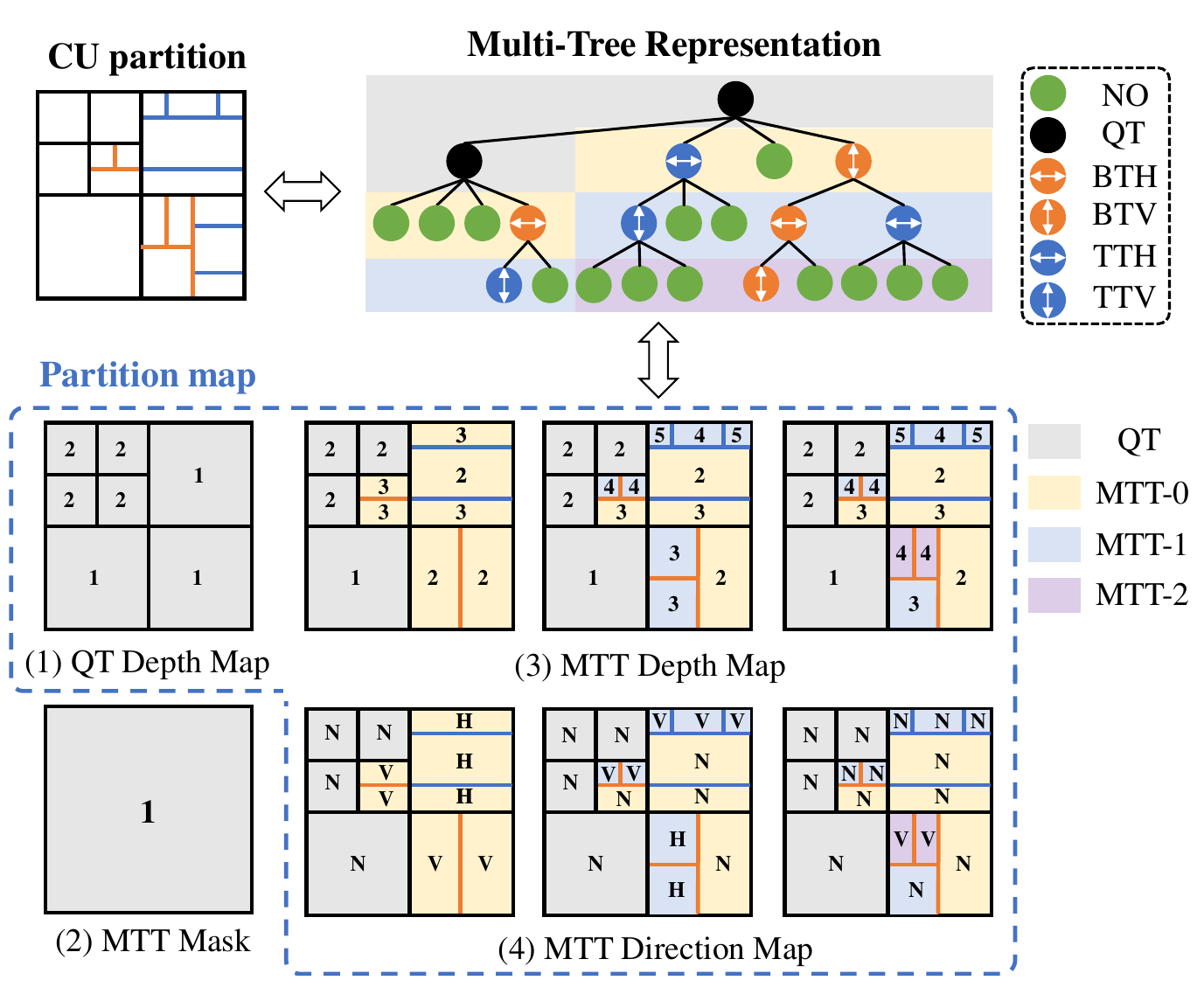}
	\caption{
 \textbf{An example of the improved partition map}. The original partition map includes three components: QT depth map, MTT depth map, and MTT direction map. It can be converted back and forth with the multi-tree representation. To represent the partition results of inter coding, we introduce an MTT mask that indicates whether early terminating MTT splits.
 } 
	\label{fig:comprasion}
\end{figure}

\begin{figure*}[t]
	\centering 
    \includegraphics[trim=0cm 0cm 0cm 0cm, clip, width=0.9\textwidth]{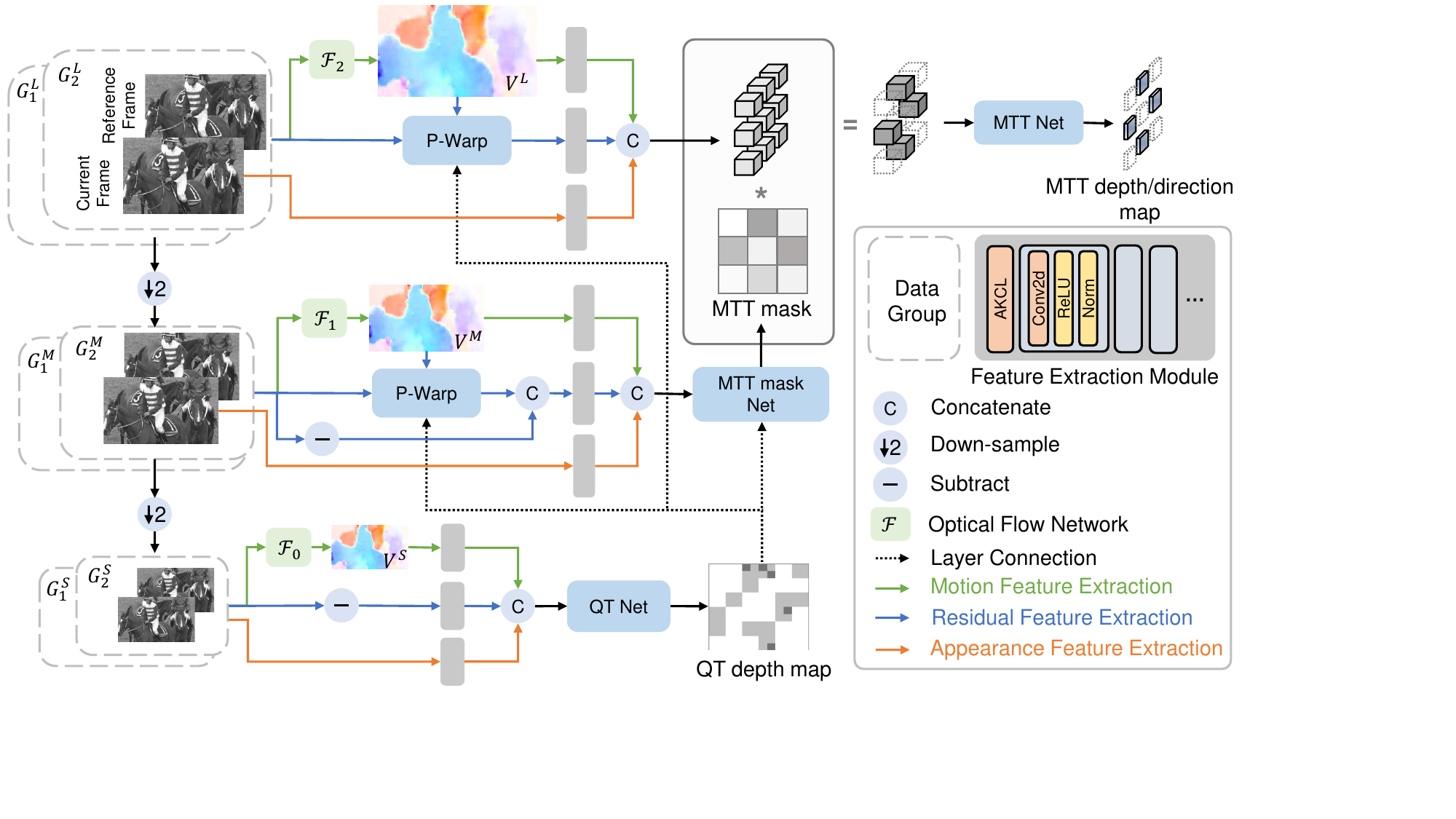}
	\caption{\textbf{Structure of the proposed neural network.} 
It takes the luma component of the current frame and its two nearest reference frames as input, and progressively generates a partition map.
The three input frames are grouped into two data groups, $G_1^L$ and $G_2^L$, which are then passed through two sequential bicubic downsampling modules to produce the other data groups, $\{G_i^M\}_{i=1}^2$ and $\{G_i^S\}_{i=1}^2$.
These groups $\{G_i^L\}_{i=1}^2$, $\{G_i^M\}_{i=1}^2$ and $\{G_i^S\}_{i=1}^2$ are utilized in a coarse-to-fine manner to predict the QT depth map, MTT mask, and MTT depth/direction map, respectively, with higher-resolution inputs corresponding to finer block partitions. Here, ``AKCL” denotes the asymmetric kernel convolution layer~\cite{shi2019ays}. 
 } 
	\label{fig:partitoin_net}
\end{figure*}

\subsection{The Improved Partition Map}

The partition map is a complete representation of the QT+MTT structure for intra coding \cite{fal2023}, as shown in Fig. \ref{fig:comprasion}. It comprises a QT depth map, three MTT depth maps, and three MTT direction maps. It can be mutually converted with the QT+MTT partition tree. Specifically, a single QT depth map represents the QT partition, where the depth value indicates the number of QT partitions\cite{fal2019hm}. MTT depth maps build upon the QT depth map by further adding depth values, corresponding to MTT nodes as child nodes of QT nodes or root nodes in partition search. When a node is split using a Binary Tree (BT), the depth increases by 1, and for a Ternary Tree (TT), the depth of the middle child node increases by 1, while the depth of the two end nodes increases by 2, considering consistency with depth and granularity of partition. However,  depth maps alone are not sufficient, thus, direction maps are designed to construct a complete representation. In particular, each layer of the MTT direction maps operates independently and aligns with the MTT depth map. When a CU is horizontally split, the corresponding value in the MTT direction map is set to 1. The value is set to -1 when a CU is vertically split. Otherwise, the value is set to 0. Thus, a partition map with four layers can be utilized to represent any QT+MTT partition tree. Note that multiple acceleration levels are achieved with the hierarchical representation, which is important in realistic scenarios.

Although the partition map provides a complete representation of the QT+MTT structure, its direct application in VVC inter coding may not be suitable. To adapt the representation to the coarse partition,  we introduce a new trade-off flag called the MTT mask, which is associated with the QT depth map and indicates early terminating MTT splits for CTUs. Specifically, when the QT nodes or root nodes stop further splitting, the MTT mask is set to false; otherwise, it is set to true. 
Thus, the improved partition map consists of the QT depth map, MTT depth/direction map, and MTT mask. 
The new representation provides two benefits over the previous one, from both the neural network architecture and post-processing perspectives.
On the one hand, it effectively models the coarse distribution of block partitioning in inter coding, thereby dynamically avoiding the redundant complexity of network inference. On the other hand, by utilizing the MTT mask, a more flexible trade-off between complexity  and coding efficiency can be achieved through an appropriate RDO acceleration strategy.

\section{Neural Network Architecture} \label{sec:Architecture}

\subsection{Overview}

The overall architecture of the proposed neural network for predicting the partition map is shown in Fig.~\ref{fig:partitoin_net}. 
The luma components of the current frame, the most recent forward reference frame, and the most recent backward reference frame are fed into the network to predict a partition map that models the block partitioning results of the current frame \footnote{If the backward reference frame is missing, the forward reference frame is replicated as the backward reference frame.}. 
The network consists of three parts: the QT depth map, MTT mask, and MTT depth/direction map prediction. Specifically, the current frame and a certain reference frame form a data group denoted as ${G}_i^L$, where $L$ means the large size and $i \in \{1, 2\}$ represents the index of the input data group. Then, $G_i^L$ is first sent through two sequential bicubic downsampling modules to obtain the other two data groups, $G_i^M$ and $G_i^S$, where $M$, and $S$ denote medium, and small sizes, respectively. 
The groups $\{G_i^L\}_{i=1}^2$, $\{G_i^M\}_{i=1}^2$, and $\{G_i^S\}_{i=1}^2$ are  used to predict the QT depth map, MTT mask, and MTT depth/direction map progressively, such that higher-resolution inputs correspond to the partitioning of smaller blocks.
To reduce computational overhead, the predicted MTT mask selects informative CTU features, which are then fed into the MTT-Net.

\subsection{QT Depth Map Prediction}
We first extract the spatial-temporal features from the data groups $\{G_i^S\}_{i=1}^2$, and then combine all the features to serve as the input to the QT Net for predicting the QT depth map.

\subsubsection{Spatial-Temporal Feature Extraction}
Referred to \cite{pan'21}, we adopt a multi-information fusion module to extract spatial-temporal features. Specifically, the current frame, the residual between the current frame and two reference frames, and the bidirectional optical flow estimated by the pretrained optical flow network \cite{tang2023offline} are fed into the feature extraction modules separately. These modules are used to obtain the appearance, residual, and motion features, respectively, denoted as $\mathbf{F}_\mathrm{app}$, $\mathbf{F}_\mathrm{res}$, and $\mathbf{F}_\mathrm{motion} \in \mathbb{R}^{\frac{H}{r} \times \frac{W}{r} \times C_f}$. Here, $H$ and $W$ denote the height and width of the input, $r$ is the downscaling factor (set to 4 in our experiments), and $C_f$ represents the number of feature channels.
The appearance feature captures texture complexity, while the residual and motion features reflect motion intensity.  Each feature extraction module consists of three convolutional layers: the first layer is an asymmetric kernel convolutional layer (AKCL) \cite{shi2019ays} with a stride of 2, designed to learn directional features from the luma component, followed by two residual blocks with strides of 2 and 1. 
Finally, the extracted features are concatenated along the channel axis to form the unified spatial-temporal feature $\mathbf{F}_\mathrm{st}$, defined as: 
\begin{equation} \mathbf{F}_\mathrm{st} = [\mathbf{F}_\mathrm{app} ; \mathbf{F}_\mathrm{res} ; \mathbf{F}_\mathrm{motion}] \in \mathbb{R}^{\frac{H}{r} \times \frac{W}{r} \times C}, \end{equation}
where ``$[;]$'' denotes channel-wise concatenation. The number of channels $C$ is empirically set to 32 in our implementation.

\begin{figure}[t]
	\centering 
    \includegraphics[trim=0cm 0cm 0cm 0cm,width=0.45\textwidth]{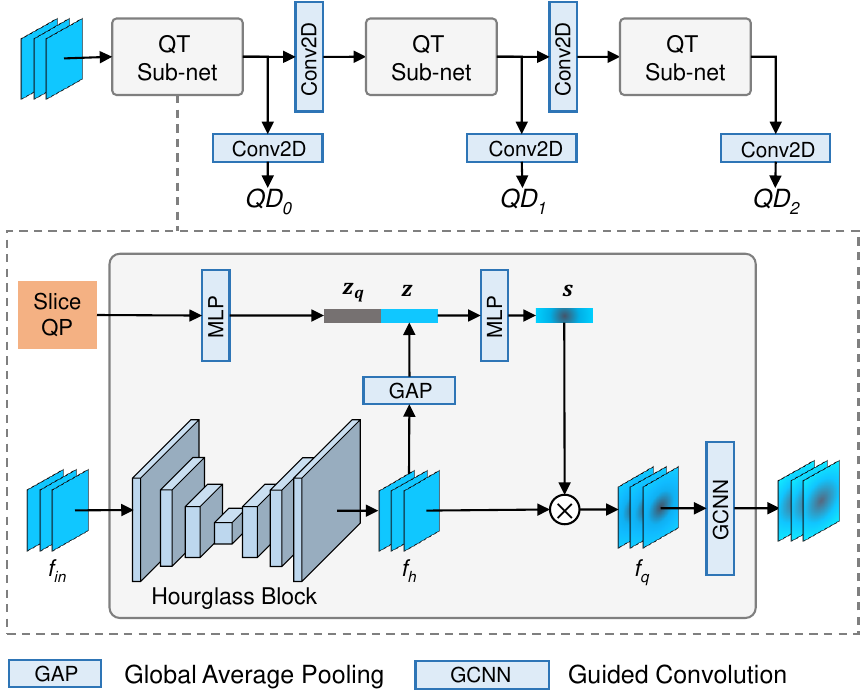}
	\caption{\textbf{Structure of the QT Net.} It consists of three sub-networks, generating three intermediate QT depth maps. Each sub-network comprises an hourglass block with a QP modulation layer and a guided convolutional layer to integrate the targeted quantization parameters and CTU boundary.} 
	\label{fig:qt_net}
\end{figure}

\subsubsection{QT Net}
The concatenated feature $\mathbf{F}_\mathrm{st}$ is fed into QT Net to predict the QT depth map, where each value represents the depth of the QT partition. A key challenge in partition map prediction is ensuring that the predicted results follow quadtree partitioning rules, referred to as \textit{local consistency}.
Previous approaches often take the CTU as input and rely on handcrafted post-processing algorithms to generate valid QT depth maps \cite{fal2023}. Inspired by human pose estimation \cite{newell2016stacked}, we propose a stacked top-down and bottom-up processing approach that encourages neural networks to develop a coherent understanding of the partition structure, thereby improving the local consistency of the predicted results in an end-to-end manner.
The architecture of QT Net, shown in Fig.~\ref{fig:qt_net}, consists of three stacked, compression-aware sub-networks, each producing an intermediate QT depth map, denoted as \(\{QD_1, QD_2, QD_3\}\in\{0,1,2,3,4\}^{\frac{H}{4}\times\frac{W}{4}}\). Each sub-network employs a 3-level hourglass structure \cite{newell2016stacked}, processing features at both local and global contexts. During training, the L1 loss between each intermediate predicted map and the ground truth is computed. At inference, only the final output \(QD_3\) is used.
By progressively aligning local predictions with their adjacent regions, the network naturally learns to generate depth maps that comply with quadtree constraints.

The structure of the proposed QT sub-network is depicted in Fig.~\ref{fig:qt_net}, including three key components: an hourglass block, a quantization parameter modulation layer, and a guided convolutional layer.
Within each sub-network, the input feature is initially processed by a 3-level hourglass block, which efficiently integrates both local and global cues through top-down and bottom-up processing, to obtain a multi-scale feature denoted as $\mathbf{F}_h$.
Next, the feature $\mathbf{F}_h$ is modulated with the QP of the current encoding frame to adapt to P/B frames with varying compression levels. Specifically, global average pooling (GAP) is applied to $\mathbf{F}_h$, producing channel-wise statistics $\mathbf{z} \in \mathbb{R}^C$. These statistics $\mathbf{z}$ are then combined with the embedding of the slice QP, denoted as $\mathbf{z}_q \in \mathbb{R}^C$, to compute the channel attention weights $\mathbf{s}$ as follows: \begin{equation} \mathbf{s} = \mathrm{Sigmoid}\left(\mathbf{W_2} \mathrm{ReLU}\left(\mathbf{W_1}[\mathbf{z}_q;\mathbf{z}]\right)\right), \end{equation} where $\mathbf{W_1} \in \mathbb{R}^{2C \times 2C}$ and $\mathbf{W_2} \in \mathbb{R}^{2C \times C}$ are learnable weight matrices. The feature $\mathbf{F}_h$ is then weighted by the attention weights $\mathbf{s}$ to obtain the modulated feature: $\mathbf{F}_q = \mathbf{F}_h \odot \mathbf{s}$.
To further enhance the spatial localization of features, we employ a guided convolutional layer (GCNN) inspired by learning-based in-loop 
 filters \cite{tianyi2019multiframe}. It utilizes a binarized grid map $\mathbf{M}$ to indicate CTU boundaries, where CTU boundaries are marked as -1, while all other regions are marked as 1:
\begin{equation}
\mathbf{M}({x}, {y}) = 
\begin{cases}
-1 & \text{if } x \text{ or } y \bmod \frac{128}{r} = 0, \\
1 & \text{otherwise},
\end{cases}
\end{equation}
where $x$ and $y$ represent the horizontal and vertical coordinates, respectively.
Then, we concatenate the grid map $\mathbf{M}$ with the modulated feature $\mathbf{F}_q$ along the channel axis. This combined feature map is then passed through a convolutional layer to produce the output feature $\mathbf{F}_{\text{out}}$ of the sub-network.

\subsection{MTT Mask Prediction}

Like in QT depth map prediction, spatial-temporal features are first extracted from the medium-sized data group $\{G_i^M\}_{i=1}^2$ and then fed into the MTT mask network to predict the probabilities of the MTT mask. Drawing inspiration from the relationship between block-based motion estimation and block partitioning in VTM, we incorporate the predicted QT partition into motion feature extraction. Specifically, during the inter coding of VVC, block-based motion compensation is typically performed under the assumption that \textit{the motion of pixels within a block tends to be uniform} \cite{li2024TBC,li2024ustc,10849917}. This assumption makes the motion field easy to represent with low complexity in a block-based coding framework.
To align with this hypothesis, we introduce partitioning-adaptive warping (P-warping), which enables partitioning-adaptive motion compensation using the predicted QT depth map. This operation empowers the MTT mask network to  capture the temporal alignment results provided by the QT Net effectively.

\begin{figure}[t]
	\centering 
     \includegraphics[trim=0cm 0cm 0cm 0cm,width=0.45\textwidth]{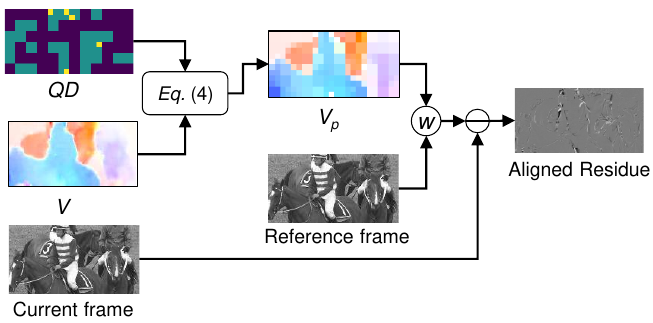}
	\caption{
 \textbf{Pipeline of the partitioning-adaptive warping (P-warping)}, which estimates block-based motion compensation using the partitioning-adaptive optical flow \( \mathbf{V}_p \), where \( QD \) represents the predicted QT depth map, \( \mathbf{V} \) denotes the optical flow, and \(\mathbin{\textcircled{w}}\) indicates the warping operation.
} 
	\label{fig:dynamic_local}
\end{figure}

The pipeline for P-warping is shown in Fig. \ref{fig:dynamic_local}. Unlike the standard warping operation, P-warping first transforms the original optical flow \(\mathbf{V}\) into the partitioning-adaptive optical flow \(\mathbf{V}_p\) using the predicted QT depth map \(QD\), as follows:
\begin{equation}
    \begin{aligned}
\label{eq:train_dlm}
\mathbf{V}_p & = \sum_{k=0}^{2} \left( (QD - \lfloor QD \rfloor) \odot \mathbf{V}^{(k+1)} \right. \\
&\quad + \left. (\lceil QD \rceil - QD) \odot \mathbf{V}^{(k)} \right) \odot \nVdash(k\leq QD<k+1),
\end{aligned}
\end{equation}
where \(\nVdash(\cdot)\) is the element-wise indicator function, which outputs a value of 1 when the condition is satisfied and 0 otherwise. The floor and ceiling operations are denoted by \(\lfloor \cdot \rfloor\) and \(\lceil \cdot \rceil\), respectively. \(\mathbf{V}^{(k)}\) represents the optical flow obtained through average pooling with a block size of \(2^{7-k}\), followed by nearest-neighbor upsampling with the same block size:
\begin{equation}
    \mathbf{V}^{(k)} = \text{upsample}\left(\text{avgpool}(\mathbf{V}, 2^{7-k}), 2^{7-k}\right),
\end{equation}
which corresponds to the optical flow at a specific granularity of CUs. 
For instance, if the predicted QT depth value is 1.2, the estimated motion vector for that block is computed as \( 0.8 \times \mathbf{V}^{(1)} + 0.2 \times \mathbf{V}^{(2)} \), where \(\mathbf{V}^{(1)}\) and \(\mathbf{V}^{(2)}\) correspond to the motion vectors for CUs of size \(64 \times 64\) and \(32 \times 32\), respectively.
Next, we use $\mathbf{V}_p$ for warping the reference frame $I_r$ to the current frame $I_c$ and then calculate the aligned residuals.  Compared with the aligned residual using $\mathbf{V}$, the aligned residual obtained from $\mathbf{V}_p$ reflects the results of the predicted QT depth maps to reduce temporal redundancy, thereby promoting the subsequent network to recognize areas that require further partitioning.

\subsection{MTT Depth/Direction Map Prediction}

As shown in Fig.~\ref{fig:partitoin_net}, spatial-temporal features are first extracted from the data group \(\{G_i^L\}_{i=1}^2\) and then divided into patches corresponding to the CTUs. The patches associated with less informative CTUs are discarded based on the predicted MTT mask, thereby reducing the inference cost of predicting MTT depth/direction maps. Lastly, the retained features are processed by MTT Net, which employs a down-up architecture, to predict the probabilities of MTT depth/direction maps.

\begin{figure}[t]
	\centering 
    \includegraphics[trim=0cm 0cm 0cm 0cm,width=0.45\textwidth]{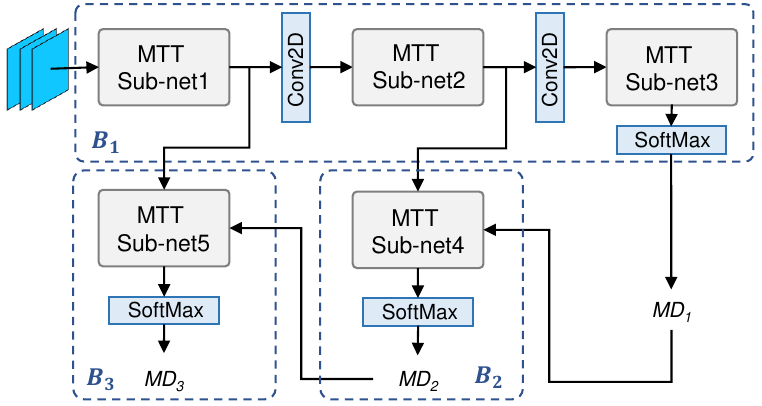}
	\caption{
\textbf{ Structure of the MTT Net.} 
It employs a similar architecture to the Down-Up CNN \cite{fal2023}, which  allows deeper outputs to be generated from shallower features. The module takes a CTU-level feature map as input and progressively outputs the probabilities of MT depth/direction maps.
} 
	\label{fig:MT_net}
\end{figure}

The structure of the MTT Net is shown in Fig.~\ref{fig:MT_net}, which adopts a similar architecture to the Down-Up CNN \cite{fal2023}. The network consists of two identical branches for predicting the MTT depth map and direction map separately. Focusing on the depth map, the network employs three branches to progressively predict three layers of the MTT depth map, namely $B_1$, $B_2$, and $B_3$. Specifically, the $B_1$ branch predicts the first layer of the MTT depth map, ${MD}_0$, using three stacked sub-networks: MTT sub-net1, sub-net2, and sub-net3. The $B_2$ branch takes the outputs from MTT sub-net2 as input to predict the second layer of the MTT depth map, ${MD}_1$, where ${MD}_0$ functions as an attention map. Similarly, the $B_3$ branch takes the outputs from MTT sub-net1 as input to predict the final layer of the MTT depth map, ${MD}_2$, where ${MD}_1$ serves as an attention map. 
The Down-Up architecture allows deeper outputs to be generated from shallower features, capturing finer details with higher granularity.

\subsection{Loss Function}

We use different loss functions to train different parts of the network. For QT depth map prediction, we utilize the L1 loss with intermediate supervision, defined as :
\begin{equation}\label{equ:qt_loss}
\mathcal{L}_{QD} = \frac{1}{3} \sum_{i=0}^2 \mathcal{L}_1(QD_{i} - \widetilde{QD}),
\end{equation}
where $QD_{i}$ denotes the QT depth map predicted by the $i$-th QT sub-net, and $\widetilde{QD}$ represents the QT depth map labels. For predicting MTT mask, MTT depth map, and MTT direction map, we employ cross-entropy loss, followed as:
\begin{align}
\mathcal{L}_{Mask} &= \mathcal{L}_{CE}(p_{M}, \widetilde{M}), \\
\mathcal{L}_{MD} &= \frac{1}{3}\sum_{n=1}^{3} \mathcal{L}_{CE}\left(p_{MD}^{(n)}, \widetilde{{MD}}_{n} - \widetilde{{MD}}_{n-1}\right), \\
\mathcal{L}_{MDir} &= \frac{1}{3}\sum_{n=1}^{3} \mathcal{L}_{CE}(p_{MDir}^{(n)}, \widetilde{{MDir}}_n).
\end{align}
Here $p_{M}$, $p_{MD}^{(n)}$, and $p_{MDir}^{(n)}$ represent the predicted probability maps for the MTT mask, depth map, and direction map at layer $n$, respectively. The corresponding ground-truth labels are $\widetilde{M}$, $\widetilde{MD}_n$, and $\widetilde{MDir}_n$.
The MTT mask label $\widetilde{M}$ is determined based on a comparison between the predicted QT depth map $Q$ and the first layer of MTT depth map $\widetilde{MD}_0$:
\begin{equation}
    \widetilde{{M}} = \begin{cases}
    0, & \text{if } Q \succeq \widetilde{MD}_0, \\
    1, & \text{otherwise},
    \end{cases}
\end{equation}
where $\succeq$ denotes that $Q(i,j) \ge \widetilde{MD}_0(i,j)$ holds for all $(i,j)$.
The total loss function is the sum of all losses, as follows:
\begin{equation}
\mathcal{L} = \mathcal{L}_{QD} + \mathcal{L}_{Mask} + \mathcal{L}_{MD} + \mathcal{L}_{MDir}.
\end{equation}

\section{Post-processing Algorithm} \label{sec:post-processing}

This section introduces a post-processing algorithm developed for the improved partition map. Similar to various existing methods that predict a CTU or large CU partition as a whole\cite{wsl2022intra, galpin2019intra, fal2019hm, fal2023}, a post-processing algorithm is necessary to generate a partition structure in accordance with the VVC standards. Moreover, we propose a dual threshold decision scheme that enables a flexible trade-off between complexity reduction and coding efficiency. 

\subsection{Map Tree-Based Post-processing Algorithm} 

Given the predicted partition map, we apply the map tree-based post-processing algorithm proposed in \cite{fal2023} to generate a set of standard-compliant split mode decisions. 
The algorithm includes two stages. In the first stage, a depth-first search order is used to retrieve all possible partition structures starting from the root node, generating a collection of candidate partition maps. 
Thus, each node maintains a temporary partition map named the \textit{map tree}. 
From the recorded map trees, multiple candidate partition paths can be derived. \textcolor{black}{In the second stage, a predefined criterion is used to select the partition path with the least error from the candidates.
Specifically, the corresponding criterion is as follows:
\begin{equation}
\begin{aligned}
    Error &= \left\|MD_t - MD_p[\text{curMTTdepth}]\right\|_1 \\
    &\quad + \left\|MDir_t - MDir_p[\text{curMTTdepth}]\right\|_1
\end{aligned}
\end{equation}
where $MD_t$ and $MDir_t$ are the temporal partition layers of the MTT depth map and direction map, respectively, and where $MD_p[\text{curMTTdepth}]$ and $MDir_p[\text{curMTTdepth}]$ are the current partition layer of the predicted MTT depth map and direction map. }
Although originally designed for CTUs with a shape of $64 \times 64$, this algorithm can be easily extended to partition maps corresponding to CTUs with a shape of $128 \times 128$ for VVC inter coding, thereby providing a set of standard-compliant partition mode decisions.

\subsection{Dual Threshold Decision Scheme} 
\begin{figure}[t]
	\centering 
    \includegraphics[trim=0cm 0cm 0cm 0cm,width=0.4\textwidth]{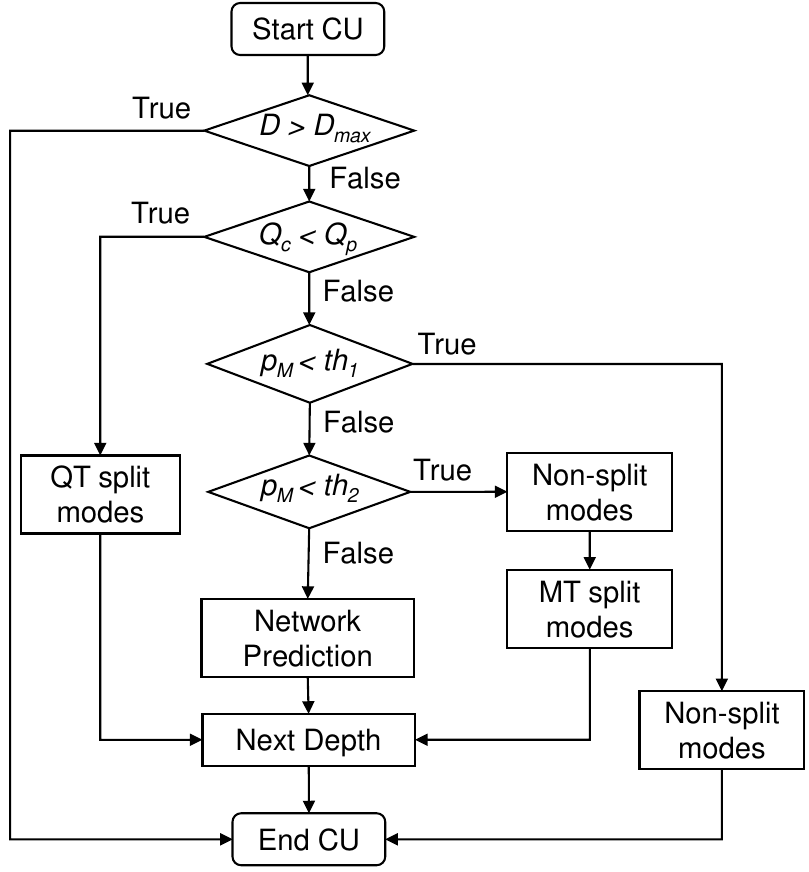}
	\caption{\textbf{Flowchart of the dual threshold decision scheme. }
When the current decision depth $D$ is not greater than the maximum depth $D_{\text{max}}$, and the current QT depth $Q_c$ is not less than the predicted QT depth $Q_p$, the dual-threshold approach is used to balance network prediction and RDO search.
$p_{M}$ denotes the predicted probabilities of the MTT mask. The “Next Depth”  indicates the progression toward deeper levels of partitioning, given a recursive pipeline of the RDO. 
 } 
	\label{fig:dual-threshold}
\end{figure}

Although the four layers of the partition map enable four levels of acceleration, a fine-grained trade-off is required between complexity reduction and coding efficiency in practical scenarios. Additionally, since neural network predictions are not always accurate, incorrect decisions can lead to significant degradation in RD performance. To address these issues, we propose a dual-threshold decision scheme that introduces a flexible trade-off between the neural network's predictions and the recursive RDO search based on the predicted MTT mask, as shown in Fig. \ref{fig:dual-threshold}. Specifically, we set a lower threshold, \( th_1 \), to early-terminate the MTT split for CTUs with predicted MTT mask Softmax values below this threshold. If the Softmax value exceeds \( th_1 \), we treat the Softmax values as a confidence score and introduce a higher threshold, \( th_2 \). For Softmax values below \( th_2 \), the CTUs perform the default RDO search; otherwise, CTUs with higher Softmax values follow the decisions of the neural network. By adjusting both thresholds, we can minimize RD performance degradation while still achieving significant encoding time savings through skipping CTUs that are difficult for the neural network. Note that the scheme does not involve QT partition acceleration, as the RD loss resulting from accelerated QT partitions is negligible in practice.

\section{Experiments}\label{sec:experimental results}

In this section, we evaluate the effectiveness of our approach. Part \ref{subsec:settings} describes dataset construction and configurations. Part \ref{subsec:performance} compares our method with relevant approaches. Part \ref{subsec:robustness} analyzes the robustness across various coding configurations. We then conduct ablation studies on key techniques in Part \ref{subsec:ablation}. Part \ref{subsec:dual_threshold} discusses threshold selection, and Part \ref{subsec:complexity} analyzes the method's complexity.

\begin{table*}
\centering
	\caption{Comparison with Related Methods at a Low Complexity Reduction Setting(\%). A Smaller BDBR Indicates Less RD Performance Degradation, While A Larger ETS Represents Greater Encoding Time Savings.}
	\label{tab:comparison_L0}
	\begin{tabular}{c|cc|cc|cc|cc|cc|cc}
		\toprule
		 \multirow{2}{*}{\textbf{Class}} & \multicolumn{2}{c|}{\textbf{\begin{tabular}[c]{c}Pan \textit{et al.}\cite{pan'21}\end{tabular}}} & \multicolumn{2}{c|}{\textbf{\begin{tabular}[c]{c}Tissier \textit{et al.} \cite{Tissier'22}\end{tabular}}} & \multicolumn{2}{c|}{\textbf{\begin{tabular}[c]{c}Peng \textit{et al.} \cite{peng2023classification}\end{tabular}}} & \multicolumn{2}{c|}{\textbf{\begin{tabular}[c]{c}Lin \textit{et al.} \cite{lin2024efficient}\end{tabular}}}  & \multicolumn{2}{c|}{\textbf{\begin{tabular}[c]{c}Ours $L_0(0,1)$\end{tabular}}} &  \multicolumn{2}{c}{\textbf{\begin{tabular}[c]{c}Ours $L_1(0.2,1)$\end{tabular}}} \\ \cmidrule(l){2-13} 

      		&  {\textbf{BDBR}}&{\textbf{ETS}} &  {\textbf{BDBR}} &  {\textbf{ETS}} &  {\textbf{BDBR}} &  {\textbf{ETS}} &  {\textbf{BDBR}}&  {\textbf{ETS}} &  {\textbf{BDBR}} &  {\textbf{ETS}}  &  {\textbf{BDBR}} &  {\textbf{ETS}}\\ \midrule

		A1                   & 2.98        & 40.97        & 1.81        & 51.10             & 2.01        & 48.10         &    2.36         &   49.75        & 1.29    & 48.32    & 1.47 & 52.78   \\
		A2                  & 4.73        & 32.92        & 1.86        & 44.60             & 1.71        & 47.20         &    2.56             &   47.66            & 2.38    & 49.86   & 2.61  &  54.53  \\
		B                  & 3.76        & 30.93        & 2.21        & 46.50             & 2.17        & 47.90         &    2.54             &   47.50           & 1.99        & 46.49   & 2.93  & 55.64   \\
		C                 & 2.56        & 25.98        & 3.20        & 43.10             & 1.98        & 44.20         &    2.67             &   52.92           & 1.39        & 38.32    & 1.52  & 39.92 \\
		E                & 2.81        & 35.16        & 1.45        & 38.70             & 1.66        & 40.80         &    1.52             &   28.67           & 1.27        & 38.94     & 2.06   & 53.61   \\
		\midrule
		\textbf{Average}        & \textbf{3.37}        & \textbf{33.19}        & \textbf{2.11}        & \textbf{44.80}             & \textbf{1.91}        & \textbf{45.64}       & \textbf{2.33} &  \textbf{45.30}  & \textbf{1.66}    & \textbf{44.39}    & \textbf{2.12}    & \textbf{51.30}     \\
		\midrule
		D               & 2.35        & 22.05        & 3.02        & 36.80             & 2.09        & 38.50        & 2.50       & 41.42   & 1.78       & 27.37  & 2.04   & 37.66 \\

		\bottomrule
	\end{tabular} 
\end{table*}

\subsection{Configuration and Settings} \label{subsec:settings}

\textbf{Dataset Construction.} \label{subsec:dataset}
We established  a large-scale training dataset for the proposed neural network. Firstly, we collected 2,672 sequences from widely recognized 4K sequence datasets, including the BVI-DVC dataset  \cite{bvi_dvc}, the Tencent Video Dataset  \cite{TVD}, and the UVG dataset  \cite{UVG}. These sequences were processed into short sequences of 32 frames across different resolutions, including 3840$\times$2160, 1920$\times$1080, 960$\times$544, and 480$\times$272. Next, we used VTM-10.0  to compress these sequences with the random access configuration defined by the \textit{encoder\_randomaccess\_vtm\_gop32.cfg} settings \cite{bossen2013common} at four basic QPs: \{22, 27, 32, 37\}, with an intraperiod of 32. Lastly, we extracted block partitioning results and converted them into partition maps. During the dataset construction, several fast tools defined in the configuration file were disabled to ensure an accurate block partitioning structure.

\textbf{Training Configuration.}
We train the neural network for a total of 1,200 epochs, which are divided into two stages. In the first stage, we progressively train the modules for QT depth map prediction, MTT mask prediction, and MTT depth/direction map prediction, requiring 500, 300, and 300 epochs, respectively. During this stage, the gradients of previously trained components are frozen while training the subsequent modules. The batch sizes for 4K sequences are set to 160, 32, and 8, corresponding to the input data groups $\{G^S_i\}_{i=1}^2$, $\{G^M_i\}_{i=1}^2$, and $\{G^L_i\}_{i=1}^2$. In the second stage, we jointly train the entire network for 100 epochs, with the batch size for the 4K sequences set to 8. During the training process, for sequences with resolutions of $1920 \times 1080$, $960 \times 544$, and $480 \times 272$, the batch sizes are 4, 16, and 64 times those used for the 4K sequences, respectively. The resolution of the input sequences is changed every 5 epochs.

All the parameters are initialized using Xavier initialization \cite{glorot2010understanding}. The Adam optimizer \cite{kingma2014adam} is used, with an initial learning rate of $1 \times 10^{-3}$ for the first stage and $1 \times 10^{-4}$ for the second stage. The learning rate is reduced by a factor of 0.98 every 10 epochs. The entire training process takes approximately two weeks to complete using eight 1080Ti GPUs. Consistent with previous work \cite{fal2023}, we trained models separately for four basic QPs: 22, 27, 32, and 37.

\textbf{Evaluation Configuration.}
We integrate the proposed method into the VVC reference software VTM-10.0 to maintain compatibility with the state-of-the-art tunable encoding complexity reduction approach for VVC inter coding \cite{Tissier'22}. Our method is evaluated using the first 65 frames from 22 video sequences, based on the JVET common test conditions \cite{Bossen2020sequence}. The evaluation uses the random access configuration defined by the \textit{encoder\_randomaccess\_vtm.cfg} setting \cite{bossen2013common}, with four basic QPs: 22, 27, 32, and 37. Additionally, we assess the method with a larger GOP size, as defined by the \textit{encoder\_randomaccess\_vtm\_gop32.cfg} setting. Note that the proposed fast algorithm does not accelerate I-frames and only speeds up the encoding of P/B-frames.
We measure encoding performance using the Bjøntegaard Delta Bit Rate (BD-rate/BDBR) metric \cite{bjontegaard2001calculation}, comparing the proposed method to the original VTM encoder. The reduction in complexity is evaluated using Encoding Time Saving (ETS), expressed as:
\begin{equation} \mathrm{ETS} = \frac{T_{\mathrm{anchor}} - T_{\mathrm{test}}}{T_{\mathrm{anchor}}} \times 100\%, \end{equation} where $T_{\mathrm{anchor}}$ denotes the encoding time of the original VTM encoder, and $T_{\mathrm{test}}$ represents the total time of the accelerated encoder, including network inference, post-processing, and encoding time. We also use Encoding Time Acceleration (ETA) to represent complexity reduction, defined as:
\begin{equation} \mathrm{ETA} = \frac{1}{1 - \mathrm{ETS}} = \frac{T_{\mathrm{anchor}}}{T_{\mathrm{test}}}, \end{equation} which reflects the nonlinear relationship between the marginal savings in encoding time and encoding efficiency. Specifically, as encoding time decreases, further reductions become increasingly difficult with similar degradation in RD performance.

All the evaluation experiments are conducted on a computer with an Intel Xeon(R) CPU E5-2690@2.60GHz and 256GB of RAM, running Microsoft Windows Server 2012 R2. Hyper-threading is disabled to reduce variability in encoding time measurements. The model is implemented using the \textit{PyTorch} platform \cite{paszke2019pytorch} without any special optimizations. GPUs are disabled during the evaluation process by default.

\subsection{Performance Evaluation}  \label{subsec:performance}
\textit{Notations for Acceleration Levels:}
The proposed method achieves scalable complexity reduction by either pruning the partition map or adjusting the thresholds of the post-processing algorithm. We define notations for different levels of acceleration. The coarse-grained acceleration level, denoted as $L_n$, refers to accelerating partition search with a pruned partition map, where $n$ ranges from 0 to 3, corresponding to four layers of the partition map. For example, $L_0$ uses the network with the lowest resolution input to obtain the predicted QT depth map, thereby skipping unnecessary QT partitions in VTM encoder. $L_1$ converts both the QT depth map and the first layer of the MTT depth/direction map into split decisions. The fine-grained acceleration level involves the dual-threshold decision scheme that balances network prediction and RDO search by adjusting the thresholds $th_1$ and $th_2$. For instance, $L_1(th_1, th_2)$ signifies that CTUs with an MTT mask prediction probability less than $th_1$ skip MTT splits, CTUs with a probability greater than or equal to $th_2$ perform MTT Net decisions, and other CTUs execute the default RDO search.

\begin{figure} 
\centering
    \includegraphics[trim=0cm 0cm 0cm 0cm,width=0.45\textwidth]{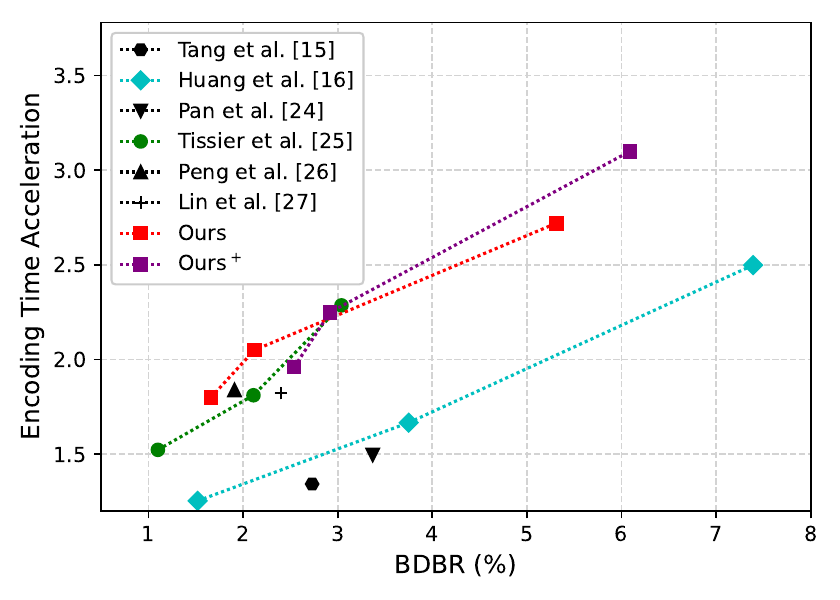}
	\caption{
\textbf{The trade-off between complexity and coding efficiency compared to other methods.} The horizontal axis represents the RD performance loss, while the vertical axis represents the encoding time acceleration ratio. The dashed line in the figure connects different trade-off points and does not indicate attainable values along the line. We present our approach at three acceleration levels: $L_0(0,1)$, $L_0(0.2,1)$, and $L_1(0.2,0.9)$. “Ours$^+$” refers to applying our method to accelerate B-frames and using the fast block partitioning method for VVC intra coding \cite{fal2023} to speed up I-frames.
} 
	\label{fig:RDTA}
\end{figure}

\textbf{Comparison with Relevant Methods.}
We compare our method with previous fast block partitioning algorithms for VVC inter coding \cite{tang'18}, \cite{huang2023precise}, \cite{pan'21}, \cite{Tissier'22}, \cite{peng2023classification}, \cite{lin2024efficient} under different complexity reduction settings. In the low complexity reduction setting, we present a detailed comparison with advanced methods \cite{pan'21}, \cite{Tissier'22}, \cite{peng2023classification}, and \cite{lin2024efficient} in Table \ref{tab:comparison_L0}. Our method achieves an average reduction in encoding time of 44.39\% to 51.30\% at two acceleration levels, $L_0(0,1)$ and $L_1(0.2,1)$, with a BDBR degradation ranging from 1.66\% to 2.12\%. Compared to the tunable encoding complexity reduction method by Tissier \textit{et al.} \cite{Tissier'22}, our method reduces encoding time by 6.5\% with a similar RD performance loss. In comparison with the state-of-the-art method by Peng \textit{et al.} \cite{peng2023classification}, which lacks scalability in complexity reduction, our approach not only reduces BDBR degradation by 0.31\% under similar complexity reduction, but also supports various levels of encoding complexity reduction. 
Moreover, our method can be extended to higher complexity reduction settings, as shown in Fig. \ref{fig:RDTA}. For example, it saves 63.21\% of encoding time with a 5.31\% increase in BDBR at $L_1(0.2,0.9)$. In contrast, few studies explore acceleration ratios beyond a 2.5x speedup while maintaining acceptable RD loss, highlighting the potential of partition map-based approaches. 

\textbf{Combination with the Fast Algorithm for Intra Coding.}
Since the proposed method does not accelerate I-frames, we further consider combining the fast algorithms for intra coding \cite{fal2023} with our method for inter coding.
The combination of both algorithms also results in different results depending on the complexity reduction settings. Augmenting the fast algorithm \cite{fal2023} upon our method does \textit{not necessarily} lead to a better trade-off between coding efficiency and complexity reduction, as demonstrated by the comparison between "Ours" and "Ours$^+$" in Fig. \ref{fig:RDTA}. At the acceleration level $L_0(0,1)$, reducing encoding time by accelerating I-frame encoding does not fully compensate for the RD loss compared to $L_0(0.2,1)$. However, at $L_1(0.2,0.9)$, where encoding time is more critical, accelerating the encoding of I-frames further reduces encoding time with modest RD performance degradation, achieving a better trade-off. Therefore, these varying performances of the combination of two types of algorithms at different settings underscore their distinct impacts on coding efficiency and complexity reduction.

\begin{table}
\caption{{Combination of fast algorithms for intra and inter coding.} The term ``Intra" refers to accelerating I-frames using \cite{fal2023}, while ``Inter" refers to the speeding up B-frames using the proposed method at acceleration level $L_0(0.2,1)$.}
\label{tab:intra_plus_inter}
\centering
\resizebox{0.35\textwidth}{!}{
\begin{tabular}{c|cc|cc}
\toprule
\multirow{2}{*}{\textbf{Class}} & \multicolumn{2}{c|}{Intra} & \multicolumn{2}{c}{Intra + Inter} \\
\cmidrule(l){2-5} 

                       & \textbf{BDBR}       & \textbf{ETS}        & \textbf{BDBR}            & \textbf{ETS}          \\ \midrule
A1  & 0.50    & 3.65  & 1.85 & 56.31          \\
A2   & 0.71   & 3.92 & 3.17  & 58.40        \\
B   & 0.69    & 3.75 & 3.53  & 59.69      \\
C   & 0.87    & 2.93 & 2.14   & 42.78   \\
E   & 2.06    & 8.40 & 3.89   & 61.02   \\
\midrule
\textbf{Average}  & \textbf{0.97} & \textbf{4.53}    &  \textbf{2.84}   & \textbf{55.64}       \\
\bottomrule
\end{tabular}
}
\end{table}

\begin{table*}[]
	\centering
	\caption{Accuracy of Network Output and Post-Processed Partition Maps (\%)}
	\label{tab:accuracy}
 
		\begin{tabular}{cc|c|c|cc|cc|cc|c}
			\toprule
			\multicolumn{1}{l}{\textbf{}}                                                   & \textbf{QP}      & \textbf{\begin{tabular}[c]{@{}c@{}}$ \mathbf{QD} $\end{tabular}} & \textbf{\begin{tabular}[c]{@{}c@{}}$ \mathbf{MTT\ mask} $\end{tabular}} & \textbf{\begin{tabular}[c]{@{}c@{}}$ \mathbf{MD_1} $\end{tabular}} & \textbf{\begin{tabular}[c]{@{}c@{}}$ \mathbf{MDir_1} $\end{tabular}} & \textbf{\begin{tabular}[c]{@{}c@{}}$ \mathbf{MD_2} $\end{tabular}} & \textbf{\begin{tabular}[c]{@{}c@{}}$ \mathbf{MDir_2} $\end{tabular}} & \textbf{\begin{tabular}[c]{@{}c@{}}$ \mathbf{MD_3} $\end{tabular}} & \textbf{\begin{tabular}[c]{@{}c@{}}$ \mathbf{MDir_3} $\end{tabular}} & \textbf{Average} \\ \midrule
   
			\multirow{5}{*}{\begin{tabular}[c]{@{}c@{}}Network\\      output\end{tabular}} 
			
                         & 22 & 63.64 & 84.49   & 63.64 & 56.10 & 45.25 & 63.22 & 46.41 & 73.71 & 61.93   \\
                         & 27 & 71.25 & 84.03   & 68.11 & 61.46 & 56.28 & 76.81 & 56.70 & 87.48 & 70.05   \\
                         & 32 & 76.36 & 84.08   & 69.23 & 64.89 & 62.09 & 81.71 & 62.64 & 91.16 & 73.83   \\
                         & 37 & 80.07 & 86.19   & 71.44 & 70.73 & 68.88 & 82.79 & 69.17 & 84.83 & 76.53   \\
			\cmidrule(l){2-11} 
			& Average & 72.83 & 84.70   & 68.10 & 63.30 & 58.13 & 76.13 & 58.73 & 84.29 & 70.58     \\

			\midrule

			\multirow{5}{*}{\begin{tabular}[c]{@{}c@{}}Post-\\      processed\end{tabular}} 
			
                       & 22 & 63.66 & 84.49   & 61.37 & 54.87 & 45.03 & 58.44 & 46.70 & 77.15 & 61.34    \\
                       & 27 & 71.11 & 84.03   & 65.40 & 59.66 & 56.06 & 73.24 & 56.62 & 87.28 & 68.95    \\
                       & 32 & 76.41 & 84.08   & 67.21 & 62.87 & 61.78 & 75.59 & 62.63 & 90.31 & 72.43    \\
                       & 37 & 80.03 & 86.19   & 68.82 & 66.19 & 68.51 & 83.90 & 69.18 & 95.26 & 77.02    \\
			\cmidrule(l){2-11}
			& Average & 72.80 & 84.70   & 65.70 & 60.90 & 57.85 & 72.79 & 58.78 & 87.50 & 69.94      \\
			\bottomrule
		\end{tabular}%
\end{table*}
Both methods exhibit some degree of overlap in RD performance degradation. From Table \ref{tab:intra_plus_inter}, we observe that the actual increase in BDBR when accelerating both I-frames and B-frames is lower than the sum of the increases when each frame type is accelerated separately. Specifically, the BDBR increase when accelerating I-frames alone is 0.97\%, when accelerating B-frames alone is 2.12\%, and when accelerating both I-frames and B-frames is 2.84\%, which is lower than the combined increase of 0.97 + 2.12 = 3.09\%. This discrepancy arises because accelerating I-frames not only affects the I-frames themselves but also impacts B-frames through inter-frame dependencies. The degradation of I-frames leads to low reference quality for subsequent B-frames, resulting in overlapping RD loss between the two types of fast algorithms.

\textbf{Accuracy of Predicted Partition Maps.}
We present the accuracy of both the network output and the post-processed partition map in Table \ref{tab:accuracy}. Post-processing, which converts the partition map into a set of standard-compliant partition mode decisions, may lead to a reduction in accuracy. On the one hand, the accuracy of the QT depth map predictions remains stable before and after post-processing, contributing to the remarkable local consistency of the predicted depth map. On the other hand, as the partition depth increases, the accuracy of the depth maps decreases, while the accuracy of the direction maps improves. This trend may be attributed to the greater proportion of zero values in the direction maps.

\begin{figure}[t]
\centering
    \includegraphics[trim=0cm 0cm 0cm 0cm, width=0.5\textwidth]{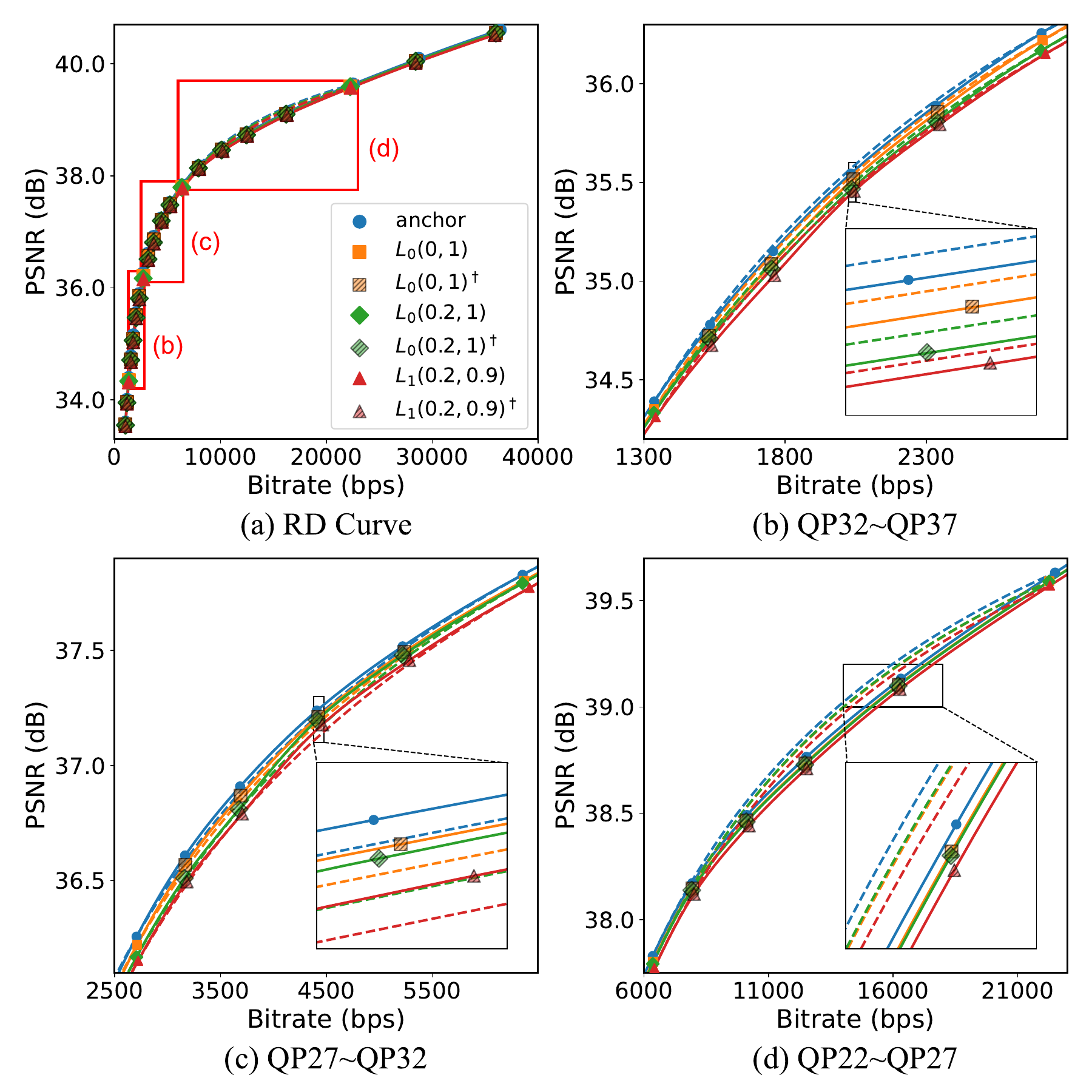}
    \caption{\textbf{Rate-Distortion Curve with Extended Basic QPs.} Taking the class B sequences as an example, we plot the RD points for basic QPs ranging from 20 to 39. The term ``anchor" refers to the original VTM encoder, while the superscript $^\dagger$ denotes the model applied to the extended QPs.  A zoomed-in view can be found in the supplementary material.
    }
    \label{fig:refine_qp}
\end{figure}

\subsection{Robustness Analysis}
\label{subsec:robustness}
Since the proposed model is trained on four basic QPs separately and the training datasets are created using VTM-10.0, it is important to evaluate the method’s performance under different encoding parameters and software versions.

\textbf{Robustness to Basic QPs.}
The proposed method supports deployment across other basic QPs, as the training dataset for each basic QP includes frames with various slice QPs, which are used as input to the QP modulation layers. In the supplementary material, we present the prediction results when the input slice QP changes, demonstrating that the QP embedding can adjust the granularity of the predicted partition structure. Building on this, we extend the model trained for each basic QP, denoted as $\{QP_i^{basic}\}_{i=1}^4$ (where $i = 1, 2, 3, 4$ corresponds to 22, 27, 32, and 37), to include the extended QPs $\{{QP_i^{basic}\pm 1, QP_i^{basic}\pm 2}\}_{i=1}^4$. Accordingly, the four pretrained models can cover all QPs ranging from 20 to 39. Figure \ref{fig:refine_qp} presents the RD curve for these QPs on the Class B sequence, where “anchor” represents the original VTM encoder, and the superscript $^\dagger$ indicates the model applied to the extended QPs. The solid and dashed lines correspond to piecewise cubic interpolation for all QPs and the four basic QPs, respectively, which form the basis for calculating BDBR. In Fig. \ref{fig:refine_qp} (b), (c), and (d), we present zoomed-in RD curves for three specific QP ranges. The results show that our approach achieves comparable RD performance on the extended QPs as on the four basic QPs across all three acceleration levels, including $L_0(0,1)$, $L_0(0.2,1)$, and $L_1(0.2,0.9)$.

\begin{table}[t]
\caption{The percentage deviation of BDBR and ETS when evaluating our approach across a total of twenty QPs, compared to the four basic QPs. A negative value of $\Delta$BDBR indicates the coding gain achieved by the extended QPs, while a positive value of $\Delta$ETS reflects the gain in encoding time savings.
}
\label{tab:qp_quantitave}
\centering
\begin{tabular}{l|c|c|c}
\toprule
& {$L_0(0,1)$} &  {$L_0(0.2,1)$} &  {$L_1(0.2,0.9)$} \\
\midrule
                       {$\Delta$BDBR$\downarrow$}   & -8.78\% & 0.37\% & -13.50\%\\ {$\Delta$ETS$\uparrow$}  & \ 0.36\% & 1.08\% & \ \ 0.24\%  \\ 
 \bottomrule
\end{tabular}
\end{table}

\begin{table*}[]
\centering

	\caption{Performance evaluation of VTM-10.0 and VTM-23.0 with a GOP size of 32.
 }
	\label{tab:latest_vtm}
	\begin{tabular}{c|cc|cc|cc|cc|cc|cc}
		\toprule

		\multirow{3}{*}{\textbf{Class}} & \multicolumn{6}{c|}{VTM-10.0} &  \multicolumn{6}{c}{VTM-23.0}\\
  \cmidrule(l){2-13} 
  & \multicolumn{2}{c|}{\textbf{\begin{tabular}[c]{c}$L_0(0,1)$\end{tabular}}} & \multicolumn{2}{c|}{\textbf{\begin{tabular}[c]{c}$L_0(0.2,1)$\end{tabular}}} & \multicolumn{2}{c|}{\textbf{\begin{tabular}[c]{c}$L_1(0.2,0.9)$\end{tabular}}} & \multicolumn{2}{c|}{\textbf{\begin{tabular}[c]{c}$L_0(0,1)$\end{tabular}}} & \multicolumn{2}{c|}{\textbf{\begin{tabular}[c]{c}$L_0(0.2,1)$\end{tabular}}} & \multicolumn{2}{c}{\textbf{\begin{tabular}[c]{c}$L_1(0.2,0.9)$\end{tabular}}}  \\ 
  \cmidrule(l){2-13} 

      		&  {\textbf{BDBR}}&{\textbf{ETS}} &  {\textbf{BDBR}} &  {\textbf{ETS}} &  {\textbf{BDBR}} &  {\textbf{ETS}} &  {\textbf{BDBR}}&  {\textbf{ETS}} &  {\textbf{BDBR}} &  {\textbf{ETS}} &  {\textbf{BDBR}}&  {\textbf{ETS}}\\ \midrule
A1      & 1.24 & 46.64 & 1.52 & 51.92 & 4.83 & 70.86 & 2.04 & 43.92 & 2.30 & 50.73 & 5.26 & 68.36 \\
A2      & 2.52 & 47.70 & 2.91 & 53.98 & 6.49 & 66.71 & 3.21 & 45.48 & 3.61 & 53.08 & 7.37 & 71.44 \\
B       & 2.09 & 45.30 & 3.02 & 54.31 & 6.76 & 70.53 & 3.10 & 45.06 & 3.97 & 54.35 & 7.03 & 69.13 \\
C       & 1.42 & 35.81 & 1.61 & 39.93 & 5.03 & 57.97 & 2.49 & 33.46 & 2.60 & 37.70 & 5.19 & 55.91 \\
E       & 1.30 & 37.37 & 2.25 & 52.07 & 4.32 & 60.90 & 1.50 & 34.65 & 2.37 & 50.17 & 3.85 & 57.61 \\
\midrule
\textbf{Average} & \textbf{1.71} & \textbf{42.56} & \textbf{2.26} & \textbf{50.44} & \textbf{5.49} & \textbf{65.39} & \textbf{2.47} & \textbf{40.52} & \textbf{2.97} & \textbf{49.21} & \textbf{5.74} & \textbf{64.49} \\
\midrule
D       & 1.74 & 24.07 & 2.19 & 28.59 & 4.78 & 36.64 & 4.60 & 21.78 & 4.97 & 23.19 & 5.49 & 35.74\\

		\bottomrule
	\end{tabular} 
\end{table*}
\begin{figure*} 
\centering
    \includegraphics[trim=0cm 0cm 0cm 0cm, width=0.8\textwidth]{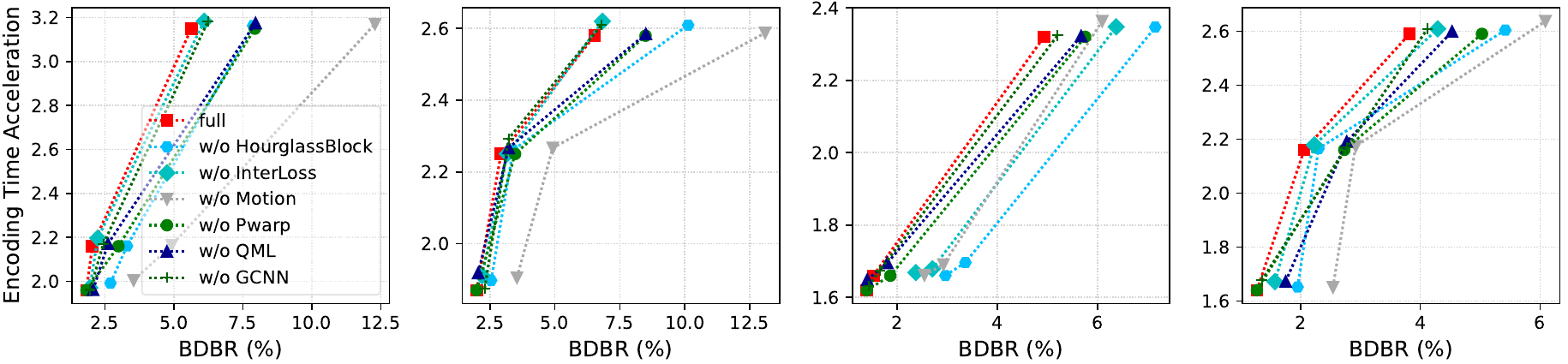}
    \caption{\textbf{Ablation Experiments on Various Techniques.} From left to right, the results correspond to Class A, B, C, and E, respectively.}
     \label{fig:ablation_import}
\end{figure*}

We further investigate the quantitative changes in BDBR and ETS by evaluating our approach across twenty basic QPs, denoted as $\mathcal{Q}_{total} = \{20, 21, 22, \dots, 39\}$, and the four basic QPs, expressed as $\mathcal{Q}_{basic} = \{22, 27, 32, 37\}$. To assess the relative differences in RD performance and complexity reduction, we use the percentage deviations $\Delta \mathrm{BDBR}$ and $\Delta \mathrm{ETS}$, defined as follows:
\begin{align}
    &\Delta \mathrm{ETS} = \left( \frac{\sum_{QP_i \in \mathcal{Q}_{total}} \mathrm{ETS}(QP_i)}{5 \times \sum_{QP_i \in \mathcal{Q}_{basic}} \mathrm{ETS}(QP_i)} - 1 \right) \times 100\%, \\
    &\Delta \mathrm{BDBR} = \left( \frac{\mathrm{BDBR}_{total}}{\mathrm{BDBR}_{basic}} - 1 \right) \times 100\%.
\end{align}
Here, $\mathrm{ETS}(QP)$ represents the encoding time savings for a given QP, and $\mathrm{BDBR}_{total}$ and $\mathrm{BDBR}_{basic}$ denote the BDBR values for $\mathcal{Q}_{total}$ and $\mathcal{Q}_{basic}$, respectively. A positive value of $\Delta\mathrm{BDBR}$ indicates that the model performs worse on the extended QPs than on the basic QPs, while a negative value of $\Delta\mathrm{ETS}$ suggests less encoding time savings on the extended QPs compared to the basic QPs. As shown in Table \ref{tab:qp_quantitave}, our approach generally achieves better RD performance and encoding time savings on $\mathcal{Q}_{total}$ compared to $\mathcal{Q}_{basic}$. 
Notably, at the acceleration level $L_1(0.2,0.9)$, our model achieves a relative coding gain of 13.50\% on the extended QPs.
These results suggest that our method can be effectively generalized to other QPs, addressing concerns about the need to train separate models for each basic QP.

\textbf{Robustness to Software Versions.}
To verify the robustness of the proposed method across different encoding configurations and software versions, we evaluate it using VTM-10.0 and VTM-23.0, with a GOP size of 32, at three different acceleration levels, as detailed in Table \ref{tab:latest_vtm}. The results show that the encoding time savings of our method remain stable, even with a larger GOP size or updated software version. When the GOP size increases from 16 to 32, our method exhibits a 0.05\% to 0.18\% increase in BDBR on VTM-10.0, due to the wider range of slice QPs and more complex dependencies across B-frames. In comparison to VTM-10.0, the BDBR increases by 0.43\% to 0.85\% on VTM-23.0, attributed to the mismatch between the training and testing datasets in terms of the software version. Considering the acceptable RD loss, the results indicate that the proposed method is robust across different software versions, owing to the similarity in core coding tools. Specifically, VTM-10.0 established the bitstream structure for VVC, with only a few encoder optimization techniques differing in higher versions.

\subsection{Ablation Experiments} \label{subsec:ablation}

\begin{figure} 
\centering
    \includegraphics[trim=0cm 0cm 0cm 0cm, width=0.4\textwidth]{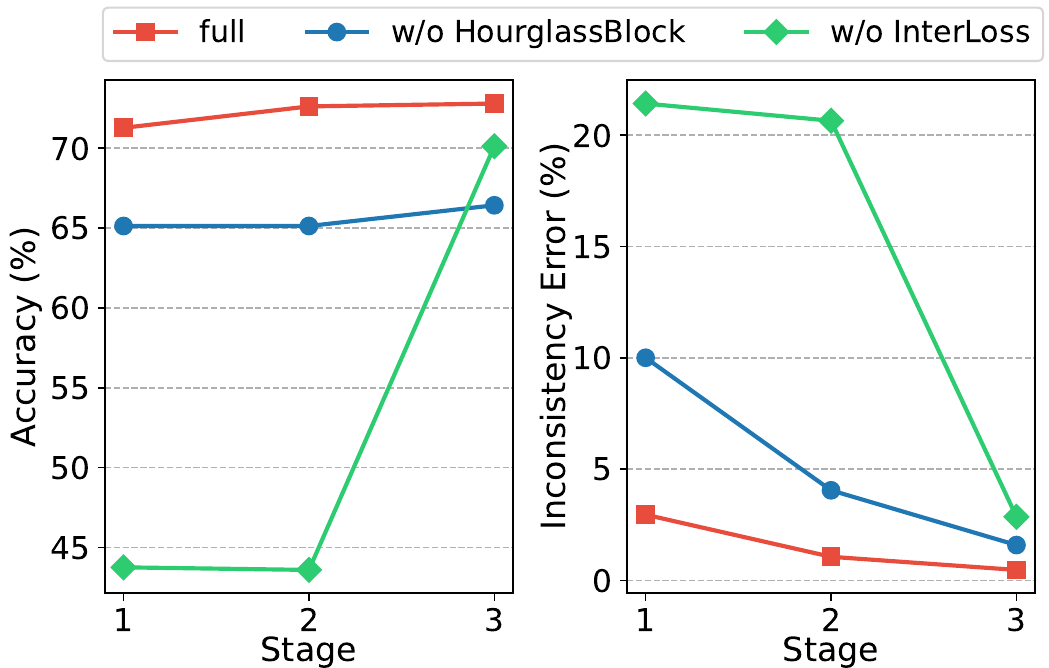}
    \caption{\textbf{The accuracy and inconsistency error across different stages of the QT-Net.} 
    The inconsistency error measures how well the prediction results conform to the quadtree partitioning rules.
Stages 1 to 3 correspond to the intermediate outputs of the three QT sub-networks.
In the \textit{w/o InterLoss} setting, where intermediate loss functions are not used to train the first two sub-networks, we employ the prediction layer from the final stage to generate the intermediate results for the first two stages.
    }
    \label{fig:refine}
\end{figure}

We conduct a series of ablation experiments to investigate the effectiveness of key techniques, \textcolor{black}{which can be divided into three categories: (1) techniques related to repeated top-down and bottom-up processing, including the hourglass block and intermediate supervision, (2) motion-related techniques, such as motion feature extraction and partitioning-adaptive warping, and (3) other techniques. The results of coding efficiency and complexity reduction when removing certain techniques are presented in Fig. \ref{fig:ablation_import}.}

\begin{figure*} 
\centering
    \includegraphics[trim=0cm 0cm 0cm 0cm, width=0.8\textwidth]{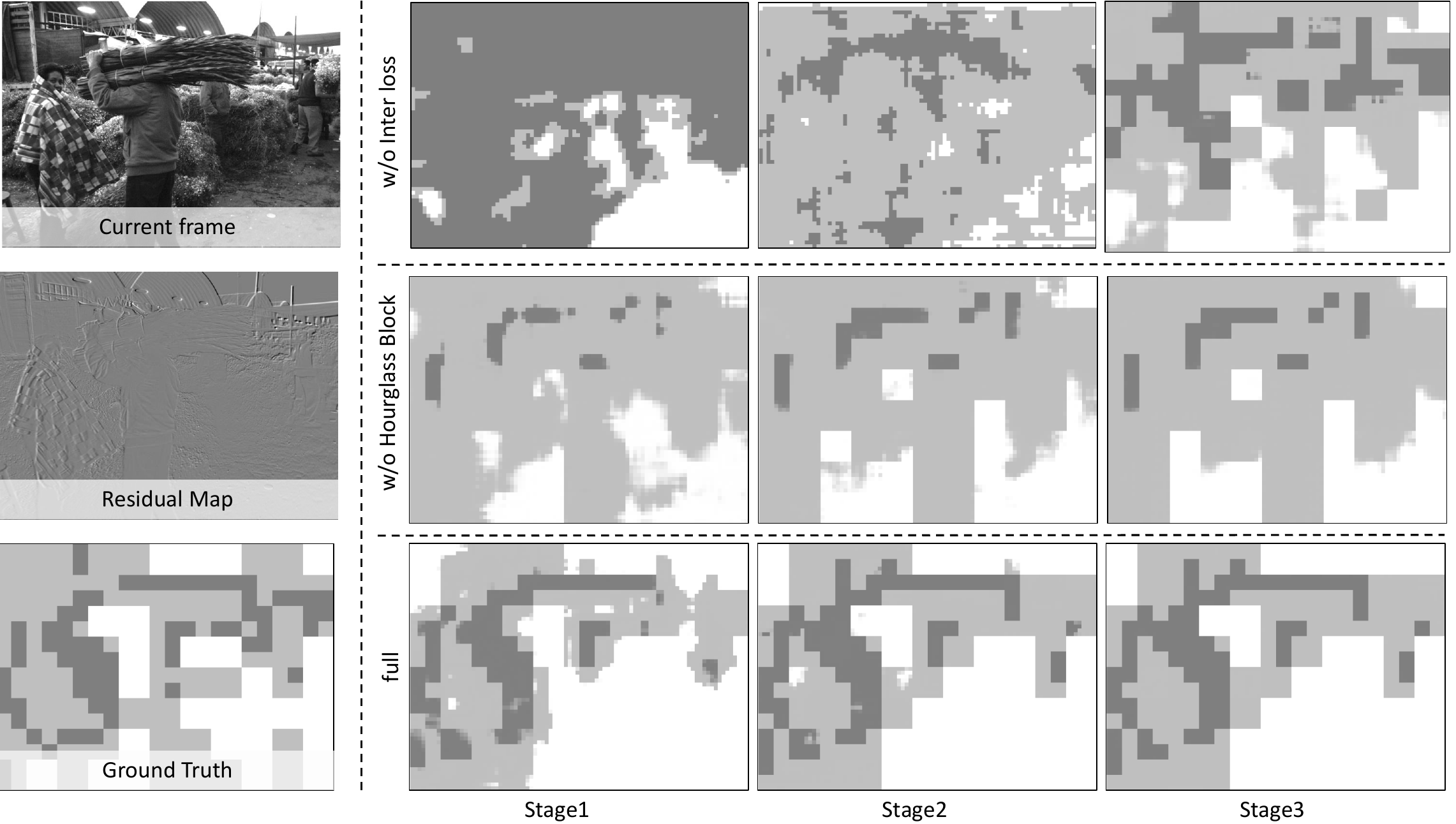}
    \caption{\textbf{Visualization of the predicted QT depth map.}
On the left, from top to bottom, are the luma component of the current frame, the residual between the current frame and the nearest reference frame, and the ground truth QT depth map, where darker colors indicate deeper partitioning.
On the right are the predictions from the three QT sub-networks under different settings. The term ``full” refers to the default configuration, while \textit{w/o HourglassBlock} and \textit{w/o InterLoss} denote the variants where hourglass blocks are replaced with dense blocks and intermediate supervision is removed, respectively.
}
    \label{fig:refine}
\end{figure*}

\textbf{Ablation Study on Repeated Top-Down and Bottom-Up Processing.}  
To enhance the accuracy and local consistency of the predicted QT depth map, we propose a progressive refinement strategy that leverages repeated top-down and bottom-up processing through stacked hourglass blocks with intermediate supervision. Specifically, the QT Net comprises three stacked QT sub-networks, each built upon hourglass modules designed to effectively capture both global and local contextual features. Intermediate supervision is applied at each stage to facilitate progressive refinement, referred to as the full setting.
To assess the contribution of each component, we conduct an ablation study with two alternative settings. In the first setting, denoted as \textit{w/o HourglassBlock}, we replace hourglass blocks with dense blocks of comparable computational complexity \cite{iandola2014densenet}. In the second setting, \textit{w/o InterLoss}, we remove the intermediate supervision defined in Eq. \ref{equ:qt_loss}, and train the model based on the final QT sub-network output. We compare the performance of all three settings in terms of prediction accuracy and inconsistency error, as shown in Fig. \ref{fig:refine}. Here, the inconsistency error quantifies the proportion of predicted values that violate quadtree partitioning rules before post-processing, thus serving as a measure of structural coherence. Moreover, Fig. \ref{fig:refine} showcases the progressive prediction results for each setting, using the second frame of the \textit{Marketplace} as an example.

Compared with the full setting, both \textit{w/o HourglassBlock} and \textit{w/o InterLoss} exhibit reduced prediction accuracy. The performance drop in \textit{w/o HourglassBlock} underscores the importance of hourglass structures in modeling multi-scale contextual features, as evidenced by the subpar partition prediction for small blocks in Fig. \ref{fig:refine}. Although \textit{w/o InterLoss} achieves comparable final accuracy, it lacks the progressive refinement, leading to inconsistent predictions. In summary, our full model not only enhances prediction accuracy but also improves structural consistency with quadtree partitioning, thereby reducing the reliance on post-processing.

\textbf{Ablation Study on Motion-Related Techniques.}
Considering the correlation between motion features and block partitioning in inter coding, we evaluate the effectiveness of motion-related techniques, including motion feature extraction and partitioning-adaptive warping. Specifically, two settings are defined as follows: \begin{itemize} \item \textit{w/o Motion} involves the removal of motion feature by setting the optical flow and residuals to constant values. \item \textit{w/o Pwarp} refers to replacing partitioning-adaptive warping with a standard warping operation. \end{itemize}

Observed from Fig. \ref{fig:ablation_import}, removing the motion branch significantly affects the trade-off between coding efficiency and encoding complexity reduction, highlighting the importance of motion features in our approach. Similarly, replacing partitioning-adaptive warping with standard warping results in a noticeable RD loss at the acceleration level $L_1(0.2,0.9)$, as the standard warping operation does not consider shallow partitioning results in motion feature extraction. We provide the aligned residuals for the two types of warping operations in the supplementary material to better reveal their differences.

\textbf{Ablation Study on Other Techniques.} 
We conduct experiments to evaluate the impact of other techniques within our approach. Each case removes a specific component, as follows: 
\begin{itemize}
    \item \textit{w/o QML} indicates a model in which the input to the QP modulation layer is set to a constant value.
    \item \textit{w/o GCNN} denotes a model that excludes the guided convolution layer from the QT sub-networks. 
\end{itemize}

The relevant results are shown in Fig. \ref{fig:ablation_import}. The RD performance is affected to varying degrees when either the QP modulation layer or the GCNN is removed, demonstrating the importance of the compression-aware capability of the QT Net.

\begin{figure}[t]
\centering
    \includegraphics[trim=0cm 0cm 0cm 0cm, width=0.45\textwidth]{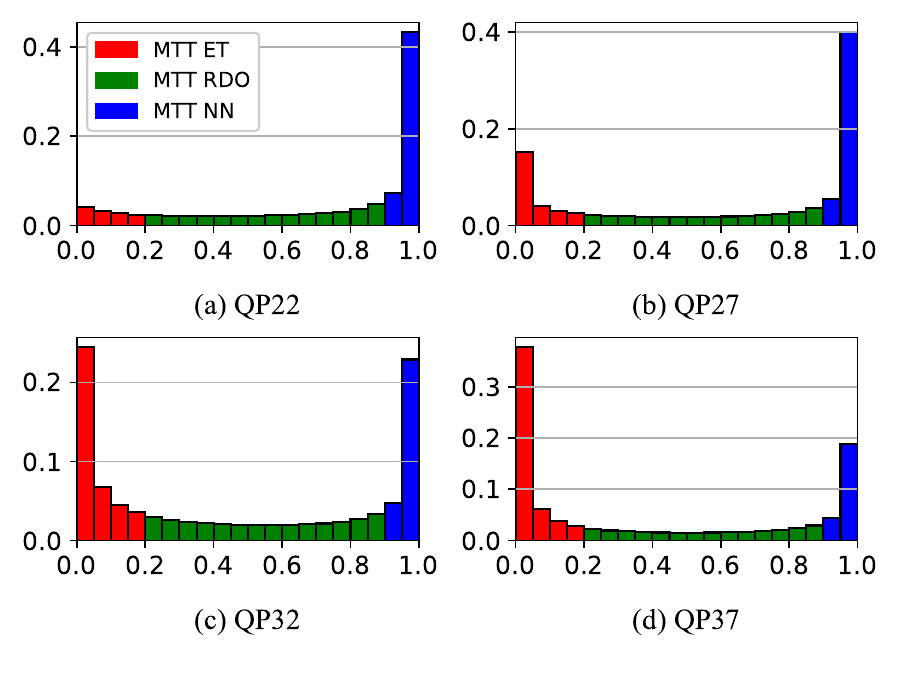}
    \caption{\textbf{Distribution of Predicted MTT Mask Values.}
The horizontal axis represents the MTT mask values, and the vertical axis denotes the distribution frequency. We set \( th_1 \) and \( th_2 \) to 0.2 and 0.9, respectively, and classify all CTUs into three types.  
``MTT ET” refers to early-terminating MTT splits, ``MTT RDO” represents the default RDO search, and ``MTT NN” denotes using the decisions predicted by the MTT Net.
    }
    \label{fig:mtt_distribution}
\end{figure}

\subsection{Analysis of Dual-Threshold Selection}
\label{subsec:dual_threshold}

Considering the significant impact of the dual-threshold decision scheme on tunable complexity reduction, it is crucial to discuss how to select the optimal threshold values. The scheme uses predicted MTT mask values, denoted as \( p_M \), which range between 0 and 1. The closer \( p_M \) is to 1, the more predominantly the corresponding CTU will be split with MTT splits. In fact, the distribution of \( p_M \) follows a bimodal pattern, as shown in Fig. \ref{fig:mtt_distribution}.  The overall CTUs can be classified into three categories—``MTT ET", ``MTT RDO", and ``MTT NN"—based on two thresholds \( th_1 \) and \( th_2 \). When \( p_M < th_1 \), MTT splits are early terminated (``MTT ET"), and when \( p_M \ge th_2 \),  MTT splits predicted by the MTT Net are applied (``MTT NN"). For the remaining CTUs, the neural network cannot confidently decide whether to split the blocks, and the default RDO search in VTM is executed (``MTT RDO"). Empirically \( th_1 \) and \( th_2 \) are set to 0.2 and 0.9, respectively,  ensuring that only a small fraction of CTUs undergo the exhaustive RDO search across various QPs, thereby achieving substantial encoding time savings with minimal RD loss.

\begin{figure}[t]
\centering
    \includegraphics[trim=0cm 0cm 0cm 0cm, width=0.4\textwidth]{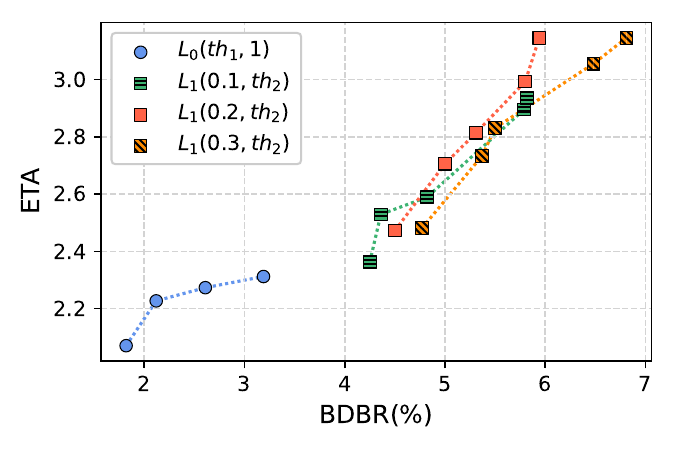}
    \caption{\textbf{Trade-off between RD performance and encoding complexity reduction with different threshold values.}
For $L_0(th_1,1)$, the threshold value $th_1$ is sequentially set to 0.1, 0.2, 0.3, and 0.4 from left to right. For $L_1(0.1, th_2)$, $L_1(0.2, th_2)$, and $L_1(0.3, th_2)$, the threshold value $th_2$ is sequentially set to 0.98, 0.95, 0.9, 0.8, and 0.7 from left to right.
    }
    \label{fig:bdbr_multi_levels}
\end{figure}

Furthermore, we validate the selected threshold values. First, we vary \( th_1 \) from 0.1 to 0.4 at the acceleration level \( L_0(th_1, 1) \), and present the trade-off between RD performance and encoding time reduction in Fig.~\ref{fig:bdbr_multi_levels}. As \( th_1 \) increases, the encoding acceleration ratio increases due to more CTUs skipping MTT split attempts, while the BDBR gradually increases. Specifically, when \( th_1 \) increases from 0.1 to 0.2, the BDBR rises by 0.3\%, and the ETA increases from 1.95 to 2.09. However, the growth rate of ETA slows down once \( th_1 \) exceeds 0.2. Therefore, we select \( L_0(0.2, 1) \) as the trade-off point, at which an average of 34.72\% of CTUs are skipped for MTT splits, as detailed in Table~\ref{tab:mtmask}, thereby reducing the MTT Net inference overhead. Next, we evaluate different combinations of \( L_1(th_1, th_2) \), with \( th_1 \) ranging from 0.1 to 0.3 and \( th_2 \) from 0.7 to 0.98. The results show that setting \( th_1 \) to 0.2 yields a higher acceleration ratio compared to other values. Thus, we set \( th_2 \) to 0.9 as the ideal threshold in our experiments.

\begin{table}[]
\caption{The ratio of CTUS early terminated by MTT mask (\%)}
\label{tab:mtmask}
\centering
\begin{tabular}{c|cccc|c}
\toprule
\multirow{2}{*}{Class} & \multicolumn{4}{c|}{QP} & \multirow{2}{*}{Average}\\
\cmidrule{2-5}
                       & 22   & 27  & 32  & 37 & \\
\cmidrule{1-6}
A       & 12.29 & 23.70 & 36.18 & 49.33 & 30.38   \\
B       & 9.35  & 20.43 & 48.47 & 52.32 & 32.64   \\
C       & 1.80  & 8.25  & 11.21 & 16.35 & 9.40    \\
E       & 39.24 & 68.46 & 76.60 & 81.57 & 66.47   \\
\cmidrule{1-6}
Average & 15.67 & 30.21 & 43.11 & 49.89 & 34.72  \\
\cmidrule{1-6}
D       & 0.00 & 0.10 & 10.24 & 9.23 & 4.89   \\
\bottomrule
\end{tabular}
\end{table}

\subsection{Complexity Analysis} \label{subsec:complexity}

We analyze the computational complexity of the proposed method, taking into account both the neural network’s inference cost and the overhead introduced by the fast algorithm.

\textbf{Inference Cost of Neural Networks.} 
To evaluate the inference cost of the neural network, we divide the proposed model into two parts: the pretrained optical flow network and the remaining components. The optical flow network serves as a plug-and-play module and can be shared across other fast inter coding algorithms \cite{peng2023classification}. For the remaining components, the QT depth map prediction requires 5.07 GFLOPs and 1.11 million parameters for a single 1080p frame. When combined with the MTT mask prediction, the total computational cost increases to 7.37 GFLOPs with 3.64 million parameters.
Compared with the previous approach \cite{Tissier'22}, which uses MobileNetV2 as the backbone and requires 3.4 million parameters, our method has a comparable number of parameters.

\begin{table*}[]
\caption{{Breakdown of time consumption at different acceleration levels, with results averaged per frame. The terms $T_\text{enc}$, $T_\text{net}$, and $T_\text{post}$ refer to the times taken by the VTM encoder, network inference, and post-processing, respectively, measured in seconds. $\rho$ indicates the proportion of the total evaluation time spent on both network inference and post-processing. GPUs are disabled by default during the testing process, except when $^+$ denotes acceleration using a single Nvidia 1080Ti.}}
\label{tab:overhead}
\centering
\begin{tabular}{c|cccc|cccc|cccc}
\toprule
\multirow{2}{*}{Class}     &    \multicolumn{4}{c|}{$L_0(0,1)$} & \multicolumn{4}{c|}{$L_0(0.2,1)$} & \multicolumn{4}{c}{$L_1(0.2,0.9)$} \\
\cmidrule{2-13}
 & $T_\text{ENC}$ &  $T_\text{NET}$ & $T_\text{POST}$ & $\rho$ (\%) &$T_\text{ENC}$ &  $T_\text{NET}$ & $T_\text{POST}$ & $\rho$ (\%) & $T_\text{ENC}$ &  $T_\text{NET}/T_\text{NET}^+$ & $T_\text{POST}$ & $\rho/\rho^+$ (\%) \\  \cmidrule{1-13}
A                                      & 1264.51                                                                                                             & 2.81                                                                                                                                                          & 0.28                                                                                                                                                       & 0.24        & 1105.43  & 8.06 & 0.29 &  0.75  & 760.78 & 30.22/2.46 & 2.09 & \ \  4.07/0.59              \\
B                                      &  \ \ 294.37                                                                                                              & 0.87                                                                                                                                                          & 0.08                                                                                                                                                      & 0.32         & \  
 \ 244.93 & 1.86 & 0.08 & 0.78  & 174.77 & 9.61/0.74 & 0.62 & \ \ 5.53/0.77               \\
C                                      &  \ \ \  \ 96.51                                                                                                               & 0.28                                                                                                                                                          & 0.02                                                                                                                                                       & 0.31         &  \ \ \ \  80.25 & 0.40 & 0.02 & 0.51 &  \  \ 59.78 & \  1.50/0.25 & 0.18 & \ \ 2.73/0.71                \\

E                                      &  \ \ \ \ 48.87                                                                                                               & 0.44                                                                                                                                                          & 0.03                                                                                                                                                       & 0.96           &  \ \ \ \  36.58 & 0.82 & 0.03 & 2.26 &  \ \ 32.42 & \ 3.21/0.37 & 0.13 & \ 9.34/1.52              \\

\cmidrule{1-13}
Average &  \ \ 426.07 & 1.10 & 0.10 & 0.46   &  \ \ 366.80 & 2.79 & 0.10 & 1.08  & 256.94 & 11.14/0.96 & 0.76 & \ \ 5.42/0.90 \\

\cmidrule{1-13}
D                                      &  \ \ \ \ 25.72                                                                                                               & 0.28                                                                                                                                                          & 0.01                                                                                                                                                       & 1.11          &  \ \ \ \  23.01 & 0.37 & 0.01 & 1.62  &  \ \ 20.37 & \ \ 0.67/0.19 & 0.03 & \  \ 3.32/1.07               \\

\bottomrule
\end{tabular}
\end{table*}

\textbf{Time Consumption.} 
we present the method overhead across different acceleration levels. To quantify the method overhead, we define \( \rho \) as follows:
\begin{equation}
    \rho = \frac{T_\mathrm{NET} + T_\mathrm{POST}}{T_\mathrm{ENC} + T_\mathrm{NET} + T_\mathrm{POST}} \times 100\%.
\end{equation}
Here, the total time consumption is decomposed into three components: VTM encoder time (\(T_\mathrm{ENC}\)), network inference time (\(T_\mathrm{NET}\)), and post-processing time (\(T_\mathrm{POST}\)). They are averaged over each frame of the JVET test sequences, without the use of GPUs to accelerate the neural networks. The results are shown in Table \ref{tab:overhead}. 
As the encoding time savings and the input resolution of the neural network increase, the method overhead also increases. For the three acceleration levels \(L_0(0,1)\), \(L_0(0.2,1)\), and \(L_0(0.2,0.9)\), the method overheads are 0.46\%, 1.08\%, and 5.42\%, respectively.
To achieve a higher acceleration ratio, we utilize a single NVIDIA 1080Ti GPU to facilitate the inference at \(L_1(0.2,0.9)\), where the inference time and method overhead are denoted by \(T_\mathrm{NET}^+\) and \(\rho^+\), respectively. Compared with inference on the CPU, the method overhead is reduced from 5.42\% to 0.90\% on the GPU. The average inference time for 4K sequences is 2.46 seconds per frame, highlighting the necessity of GPU parallel computing to achieve significant encoding time reduction.

\section{Conclusion} \label{sec:conclusion}
This paper proposes a partition map-based fast block partitioning approach for VVC inter coding. The method is developed from three key perspectives: representation, neural network architecture, and post-processing algorithms. By introducing an MTT mask, we design a novel neural network that predicts the partition map in a coarse-to-fine manner and introduces a dual-threshold decision scheme to balance neural network predictions with recursive RDO search. Inspired by the partitioning strategies of inter coding, we design several novel modules to enhance the accuracy and local consistency of the prediction results.
The experimental results demonstrate that the proposed approach achieves superior complexity reduction and BDBR performance compared to existing methods.

\textit{Limitations:} 
(1) Due to the high computational overhead of optical flow neural networks, applying the proposed method to resource-constrained devices is still challenging.
(2) Since we adopt an out-of-loop implementation, the fast algorithm cannot account for the encoding context, limiting the model's ability to adapt to various encoding environments.

In future works, we plan to reduce computational complexity, integrate the fast algorithm into the encoding loop, and extend this work to other inter modes, e.g. LDP and LDB.

\bibliographystyle{IEEEtran}
\bibliography{reference}

\clearpage


\twocolumn[
\begin{center}
    \Huge \emph{Supplementary Material for} Partition Map-Based Fast Block Partitioning for VVC Inter Coding
    \vspace{15mm}
\end{center}
]

\appendix

\section{Overview}
This supplementary material provides additional details of our proposed fast block partitioning algorithm. The remainder of the supplementary material is divided into three parts. Section \ref{sec:network} gives the detailed network architecture. Section \ref{sec:training}  describes the training strategy in detail.  Section \ref{sec:evaluating} gives more information about the inference process.

\section{Detailed Network Architecture}
\label{sec:network}

We adopt a three-level spatial pyramid network to progressively predict the partition map, enabling higher-resolution inputs to capture finer-grained block partitions. Specifically, the luma components of the current frame and its two nearest reference frames are downsampled to form a spatial pyramid \(\{G_i^L, G_i^M, G_i^S\}_{i=1}^{2}\), where the superscripts denote different spatial scales: large (L), medium (M), and small (S). These scales correspond to the prediction of the QT depth map, the MTT mask, and the MTT depth/direction map, respectively. The detailed network designs for each layer of the partition map will be described in the following sections.

\subsection{Residual Block}
\begin{figure}[t]
	\centering 
    \includegraphics[trim=0cm 0cm 0cm 0cm, clip, width=0.5\textwidth]{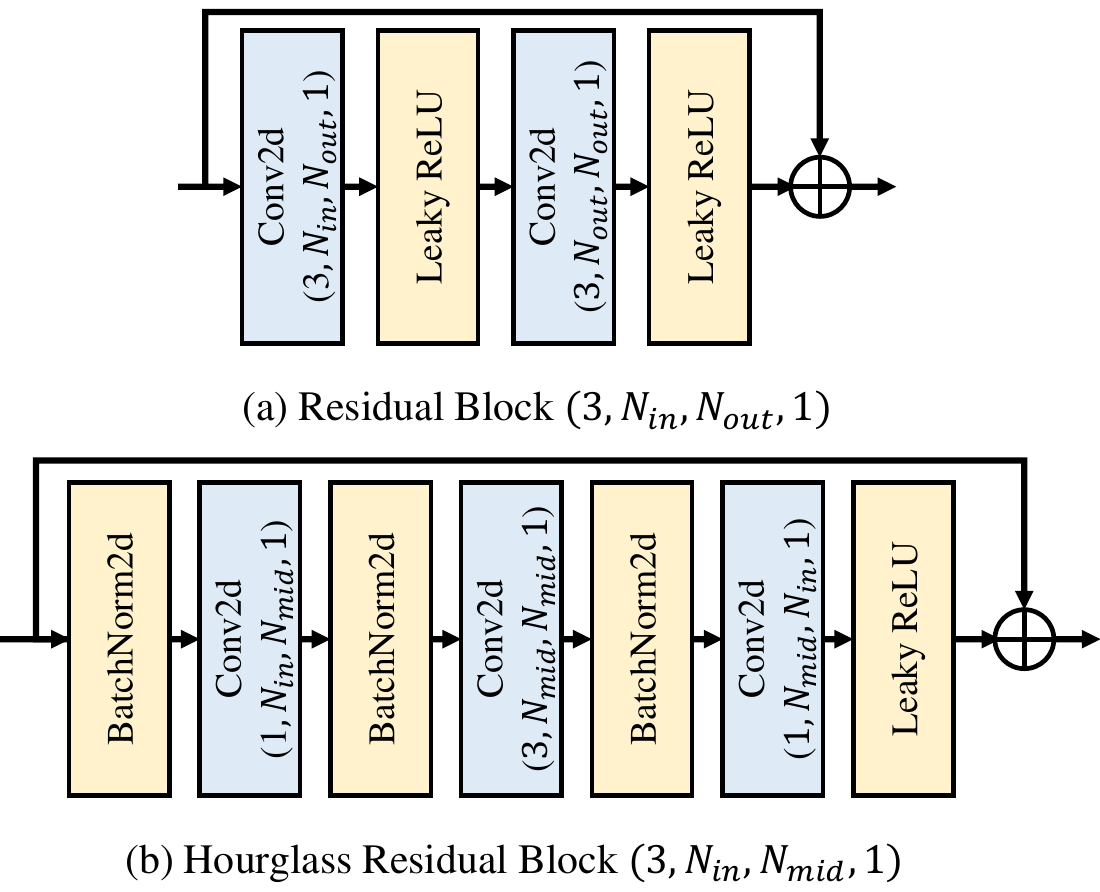}
	\caption{\textbf{Architecture of Two Types of Residual Blocks: (a)}  In the first type of residual block, the numbers $(3,N_{in},N_{out},1)$ represent the kernel size 3, the number of input channels $N_{in}$, the number of output channels $N_{out}$, and the stride 1. \textbf{(b)} The second type of residual block does not alter the dimensions of the features, and is used in hourglass block. $N_{mid}$ refers to the dimension of the internal features. The symbol $\oplus$ denotes element-wise addition.
 } 
 \label{fig:residual}
\end{figure}

The residual blocks in our implementation are categorized into two types: the regular residual block and the hourglass residual block, as shown in Fig. \ref{fig:residual}. The regular residual block is employed to adjust the spatial resolution and the number of channels. For instance, \((3, N_{\text{in}}, N_{\text{out}}, 1)\) denotes a regular residual block with a kernel size of \(3 \times 3\), input channels \(N_{\text{in}}\), output channels \(N_{\text{out}}\), and a stride of 1. In our design, this block does not include batch normalization layers  \cite{ioffe2015batch}.
The hourglass residual block, on the other hand, comprises two \(1 \times 1\) convolutional layers, one \(3 \times 3\) convolutional layer, three batch normalization layers, and a leaky ReLU activation function. It is only used in the hourglass block to extract deep features. We denote it as \((3, N_{\text{in}}, N_{\text{mid}}, 1)\), where \(N_{\text{mid}}\) indicates the number of channels in the intermediate features.

\subsection{Asymmetric Convolutio Block}
\begin{figure}[t]
	\centering 
    \includegraphics[trim=0cm 0cm 0cm 0cm, clip, width=0.45\textwidth]{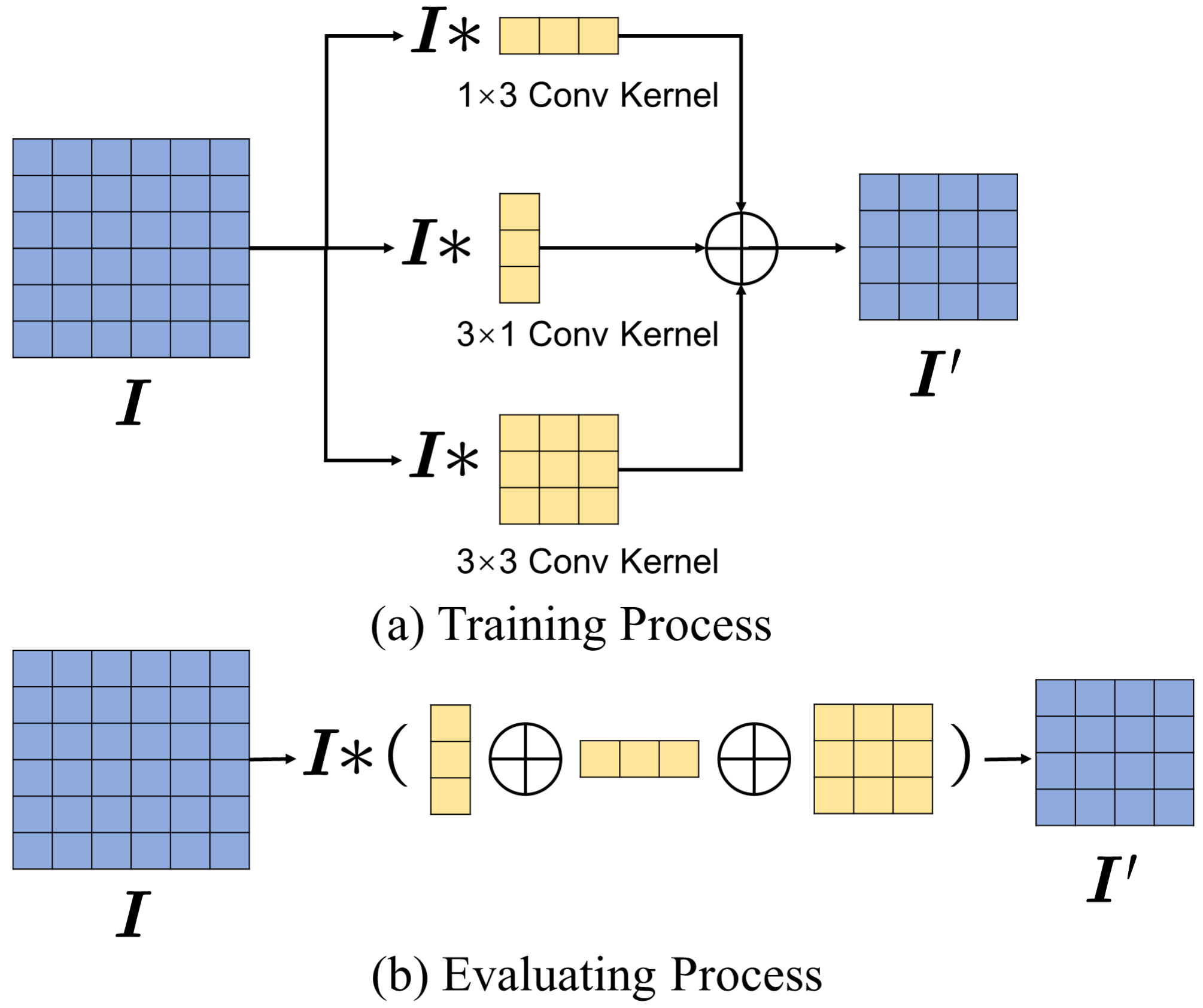}
	\caption{ \textbf{Architecture of Asymmetric Convolution Block.} 
 } 
 \label{fig:asyconv}
\end{figure}

The asymmetric convolution block employs asymmetric kernels to enhance the square convolution kernels and improve the model's robustness to rotational deformations \cite{ding2019Asymmtric}, and it has been widely applied to previous fast block partitioning algorithms \cite{shi2019ays, fal2023}. In our implementation, the asymmetric convolution kernel consists of three convolution kernels: \(1 \times 3\), \(3 \times 1\), and \(3 \times 3\), as shown in Fig. \ref{fig:asyconv}. 
During training, these kernels are optimized independently; however, during inference, they are equivalent to a single \(3 \times 3\) convolution kernel, thereby incurring no additional computational cost.

\begin{figure*}
	\centering 
    \includegraphics[trim=0cm 0cm 0cm 0cm, clip, width=\textwidth]{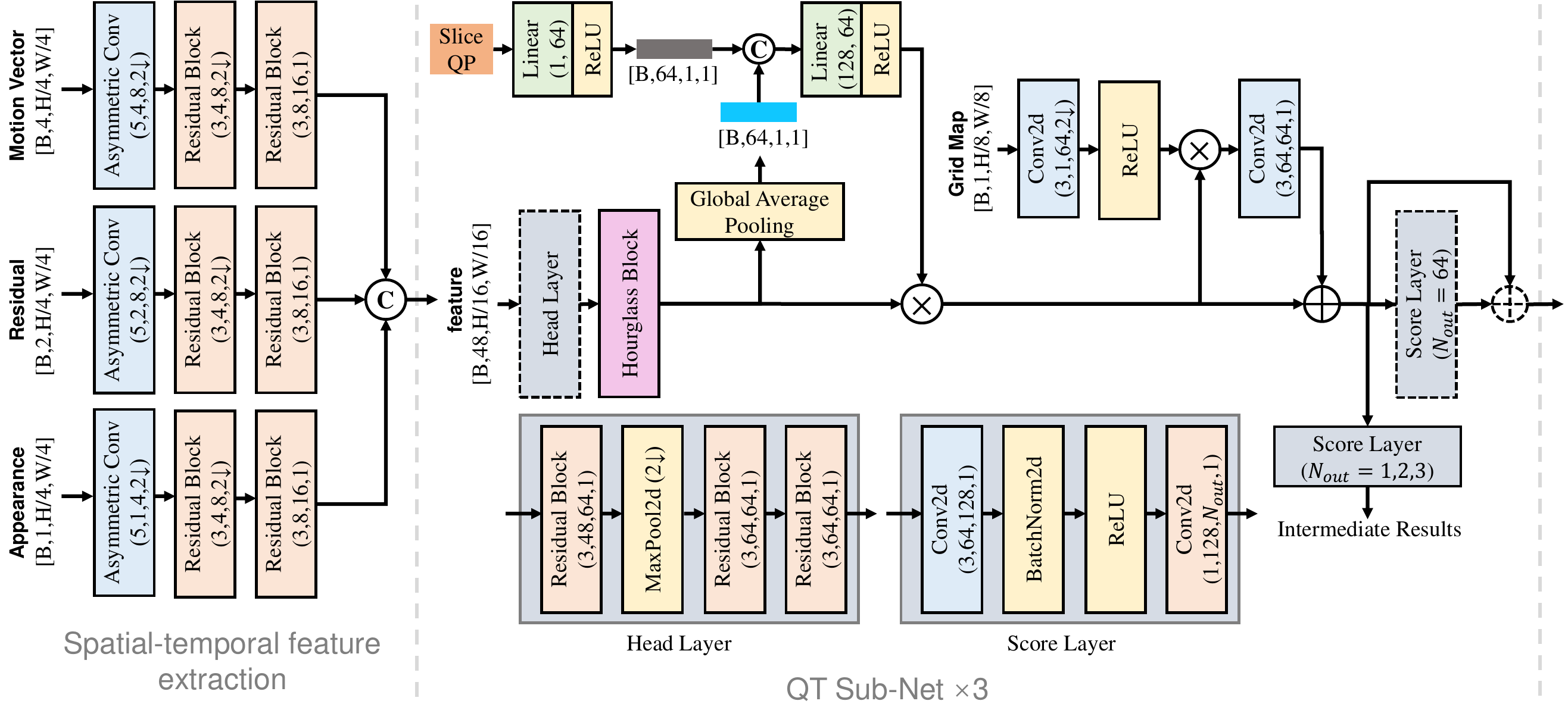}
	\caption{  \textbf{Pipeline of the QT Depth Map Prediction.} It consists of two components: spatial-temporal feature extraction and three QT sub-networks. 
 The dimensions of the features are denoted as \([B, C, H, W]\), where \(B\) corresponds to the batch size, \(C\) to the number of channels, \(H\) to the height, and \(W\) to the width.
 The dashed box around the head layer indicates its inclusion only in the first sub-net, while the dashed box around the score layer indicates its exclusion from the final sub-net. 
 } 
\label{fig:qt_branch}
\end{figure*}
\begin{figure*}
	\centering 
    \includegraphics[trim=0cm 0cm 0cm 0cm, clip, width=\textwidth]{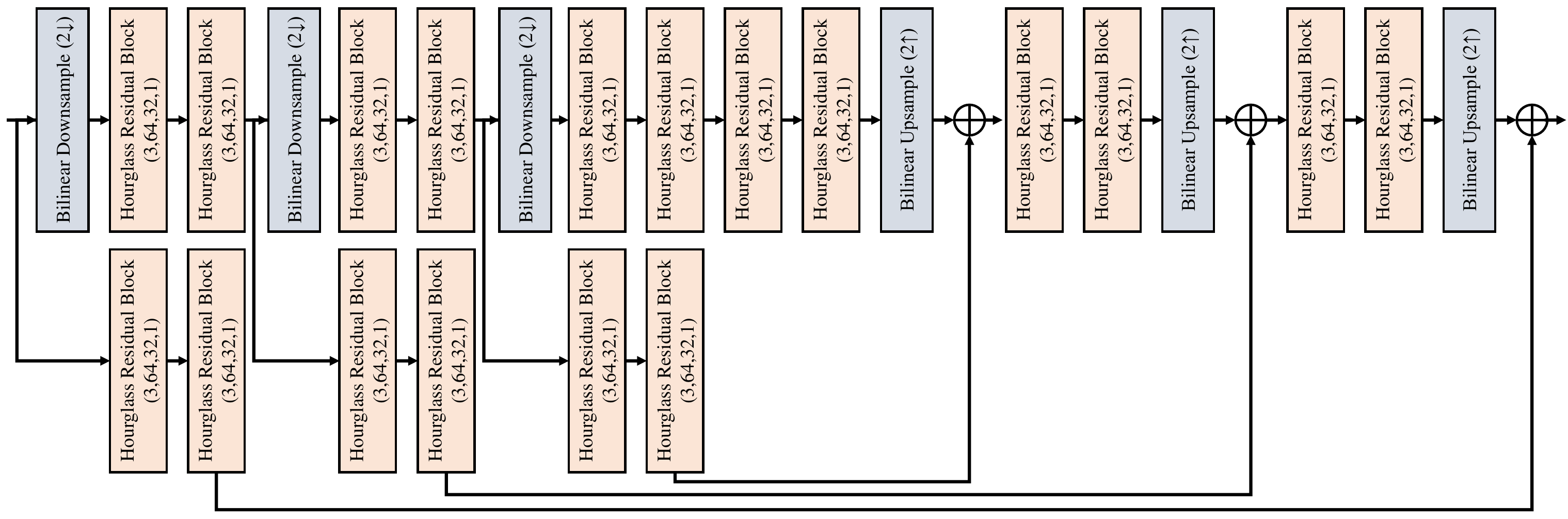}
	\caption{ \textbf{Architecture of the 3-level Hourglass Block.}
 } 
 \label{fig:hourglass}
\end{figure*}

\subsection{QT Depth Map Prediction}

We present the QT depth map prediction pipeline in Fig.~\ref{fig:qt_branch}, which consists of two main modules: spatial-temporal feature extraction and three stacked QT sub-networks. The feature extractor includes three branches to capture appearance, residual, and motion features. Before extraction, the horizontal and vertical components of the optical flow are normalized by dividing by the frame’s width and height, respectively.
The resulting features are concatenated along the channel axis. The concatenated features are first passed through a head layer to fuse different feature types. Then, three stacked QT sub-networks process the features sequentially. Each sub-network contains an optional head layer, a 3-level hourglass block (Fig.~\ref{fig:hourglass}), a slice QP modulation layer, and a guided convolution layer.
The hourglass block captures both global and local contexts in a bottom-up and top-down manner. In the hourglass block, features are downsampled via convolution and bilinear interpolation, then upsampled and merged using element-wise addition. Stacking such structures improves prediction accuracy and maintains local consistency in the QT depth map.

\begin{figure*}
	\centering 
    \includegraphics[trim=0cm 0cm 0cm 0cm, clip, width=\textwidth]{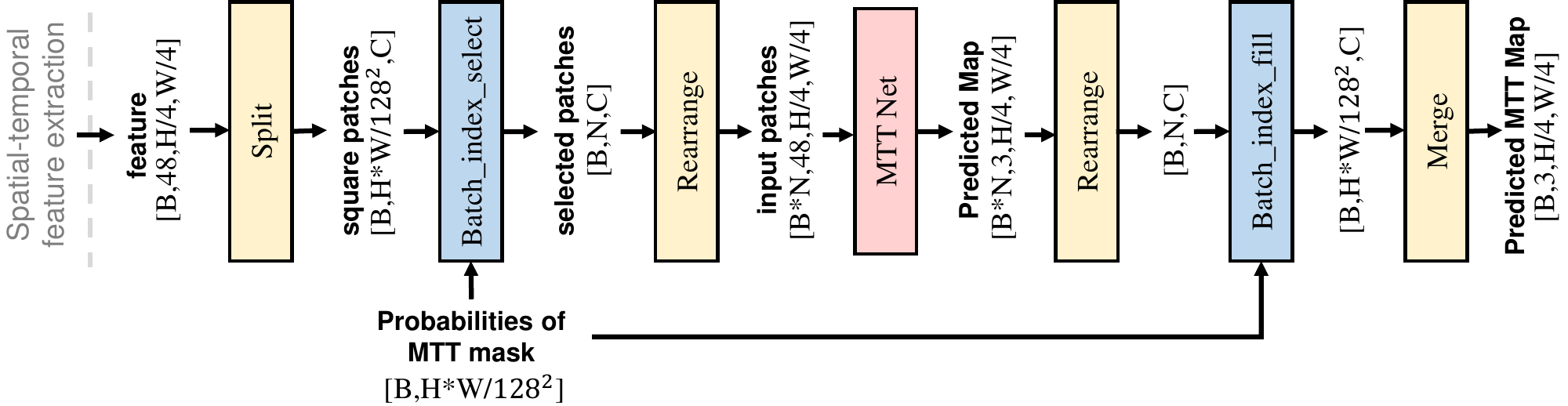}
	\caption{ \textbf{Pipeline of the MTT Depth/Direction Map Prediction.} 
 } 
 \label{fig:mtt_pipeline}
\end{figure*}

\begin{figure*}
	\centering 
    \includegraphics[trim=0cm 0cm 0cm 0cm, clip, width=\textwidth]{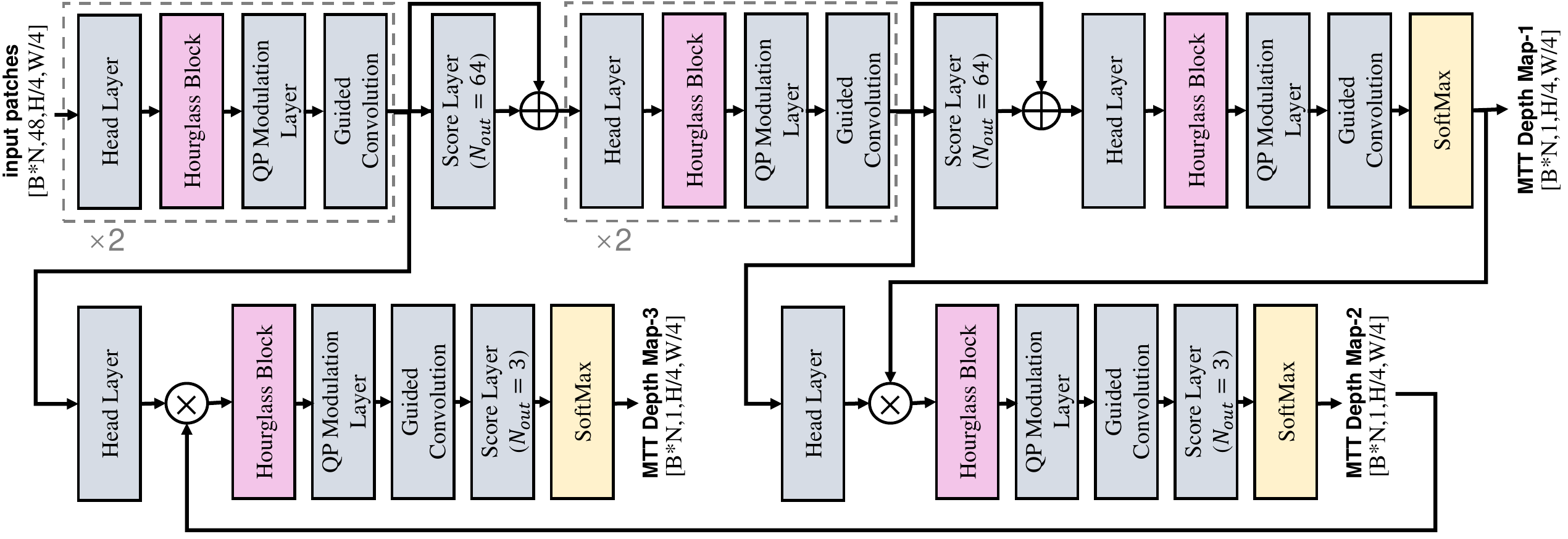}
	\caption{ \textbf{Architecture of the MTT Net.}
 } 
 \label{fig:mtt_net}
\end{figure*}

\subsection{MTT Mask Prediction}
For MTT mask prediction, there are four key differences compared to QT depth map prediction: (1) The input image resolution is downsampled by a factor of two from the original resolution, (2) the conventional inter-frame residual is replaced with the warped residual, obtained through partitioning-adaptive warping, (3) the MTT mask network consists of two hourglass blocks, rather than three, (4) each branch of spatial-temporal feature extraction is composed of an asymmetric convolution with a stride of 2, followed by three residual blocks, each with a stride of 2. The head layer consists of three Residual Blocks and three MaxPool2d layers. In the score layer, the stride of the first convolution is set to 2, and the number of output channels ($N_{\text{out}}$) is also set to 2. All other configurations remain the same as those used for QT depth map prediction, as shown in Fig. \ref{fig:qt_branch}.

\begin{figure}[htbp]
\centering
\begin{lstlisting}[language=Python]
# Pytorch-style Implementation of both batch_index_select and batch_index_fill
def batch_index_select(x, idx):
    B, N, C = x.size()
    N_new = idx.size(1)
    offset = torch.arange(B, dtype=torch.long, device=x.device).view(B, 1) * N
    idx = idx + offset
    out = x.reshape(B * N, C)[idx.reshape(-1)].reshape(B, N_new, C)
    return out
    
def batch_index_fill(x, x1, x2, idx1, idx2):
    B, N, C = x.size()
    B, N1, C = x1.size()
    B, N2, C = x2.size()
    offset = torch.arange(B, dtype=torch.long, device=x.device).view(B, 1)
    idx1 = idx1 + offset * N
    idx2 = idx2 + offset * N
    x = x.reshape(B * N, C)
    x[idx1.reshape(-1)] = x1.reshape(B * N1, C)
    x[idx2.reshape(-1)] = x2.reshape(B * N2, C)
    x = x.reshape(B, N, C)
    return x
\end{lstlisting}
\caption{\textbf{PyTorch-style pseudo code of $\mathrm{batch\_index\_select}$ and $\mathrm{batch\_index\_fill}$.}}
\label{fig:code_batch}
\end{figure}

\begin{figure*}[htbp]
\centering
\begin{lstlisting}[language=Python]
# Pytorch-style Implementation of partitioning-adaptive warping
def flow_warp(tensorInput, tensorFlow):
    if str(tensorFlow.size()) not in Backward_tensorGrid_cpu:
        tensorHorizontal = torch.linspace(-1.0, 1.0, tensorFlow.size(3)).view(1, 1, 1, tensorFlow.size(3)).expand(tensorFlow.size(0), -1, tensorFlow.size(2), -1)
        tensorVertical = torch.linspace(-1.0, 1.0, tensorFlow.size(2)).view(1, 1, tensorFlow.size(2), 1).expand(tensorFlow.size(0), -1, -1, tensorFlow.size(3))
        Backward_tensorGrid_cpu[str(tensorFlow.size())] = torch.cat([tensorHorizontal, tensorVertical], 1).cpu()
    tensorFlow = torch.cat([tensorFlow[:, 0:1, :, :] / ((tensorInput.size(3) - 1.0) / 2.0), tensorFlow[:, 1:2, :, :] / ((tensorInput.size(2) - 1.0) / 2.0)], 1)
    grid = (Backward_tensorGrid_cpu[str(tensorFlow.size())] + tensorFlow)
    return torch.nn.functional.grid_sample(input=tensorInput, grid=grid.permute(0, 2, 3, 1), mode='bilinear', padding_mode='border') 
                                           
def P_warping(depth_map, flow, refer_frame, current_frame):
    p_flow = 0
    for internal in [0, 1, 2]:
        down_stride, up_stride = 2 ** (6 - internal), 2 ** (5 - internal)
        mask = (depth_map.round() >= internal) * (depth_map.round() < (internal + 1))
        depth_1 = F.interpolate((internal + 1 - depth_map) * mask, scale_factor=8, mode='nearest') 
        depth_2 = F.interpolate((depth_map - internal) * mask, scale_factor=8, mode='nearest')
        p_flow += F.interpolate(F.avg_pool2d(flow, kernel_size=down_stride, stride=down_stride), scale_factor=down_stride, mode='nearest') * depth_1
        p_flow += F.interpolate(F.avg_pool2d(flow, kernel_size=up_stride, stride=up_stride), scale_factor=up_stride, mode='nearest') * depth_2
    aligned_frame = flow_warp(im=refer_frame, flow=p_flow)
    res = current_frame - aligned_frame
    return res
\end{lstlisting}
\caption{\textbf{PyTorch-style Implementation of Partitioning-Adaptive Warping.} 
The function \(\mathrm{P\_{warping}}\) takes the predicted QT depth map \(\mathrm{depth\_map}\), optical flow \(\mathrm{flow}\), reference frame \(\mathrm{refer\_frame}\), and current frame \(\mathrm{current\_frame}\) as inputs and outputs the warped residual. For simplicity, we describe the computation process for a single reference frame; the extension to bidirectional reference frames is straightforward.
}
\label{fig:code_pwarp}
\end{figure*}

\begin{figure*}

	\centering 
    \includegraphics[trim=0cm 0cm 0cm 0cm, clip, width=\textwidth]{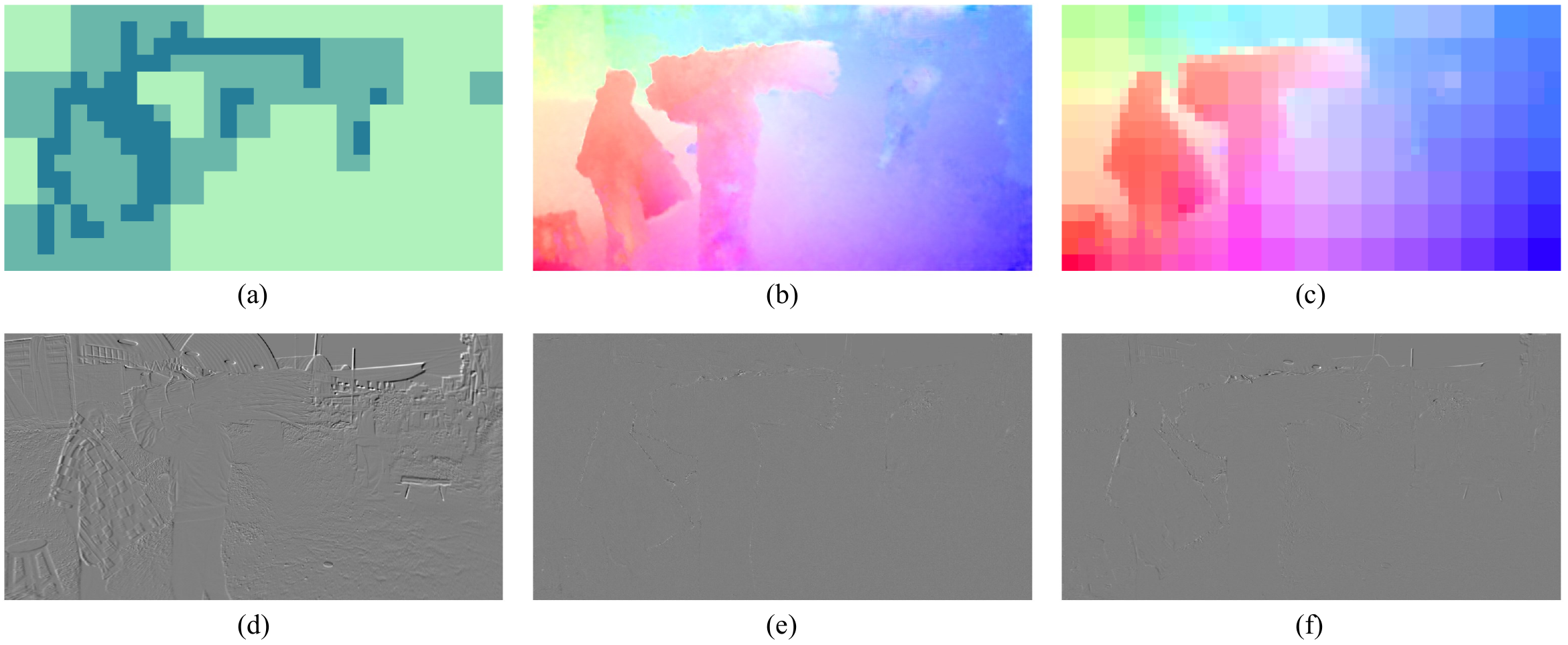}
	\caption{\textbf{Visualization of the partitioning-adaptive warping.} We select the \textit{Marketplace} video sequence from the JVET CTC \cite{bossen2013common} as an example. Subfigure (a) shows the predicted QT depth map without rounding or post-processing, where darker regions indicate higher predicted values and lighter regions indicate lower values. Subfigure (b) displays the predicted optical flow generated by SpyNet, pre-trained on motion vectors \cite{Ranjan_2017_CVPR_spynet, tang2023offline}. Subfigure (c) illustrates the partitioning-adaptive optical flow, which closely resembles block-based motion vector fields. Subfigures (d), (e), and (f) show the inter-frame residual, the warping residual based on (b), and the warping residual based on (c), respectively.
 } 
\label{fig:viz_pwarp}
\end{figure*}

\subsection{MTT Depth/Direction Map Prediction}
The detailed pipeline for MTT depth and direction map prediction is shown in Fig. \ref{fig:mtt_pipeline}. Specifically, we first divide the input features into non-overlapping square patches, each corresponding to a single CTU. The predicted probabilities of the MTT mask are then used to select the most informative patches via the $\mathrm{batch\_index\_select}$ function. The selected patches are concatenated along the batch axis and fed into the MTT Net to predict the MTT depth and direction maps. The structure of the MTT Net is depicted in Fig. \ref{fig:mtt_net}.
Finally, we use the $\mathrm{batch\_index\_fill}$ function to reconstruct the CTU-level maps into frame-level maps. The PyTorch-like implementations of the $\mathrm{batch\_index\_select}$ and $\mathrm{batch\_index\_fill}$ functions are detailed in Fig. \ref{fig:code_batch}. These functions are used to select the top-k most informative tokens and reconstruct the maps to their original size.

\subsection{Partitioning-adaptive warping}

Partitioning-adaptive warping is employed to simulate the coupling between motion estimation and block partitioning. Specifically, we use the predicted QT depth map to refine the pixel-based optical flow \cite{Ranjan_2017_CVPR_spynet, tang2023offline}, thereby generating block-based motion vector fields. These motion vectors are then used to align the reference frame with the current frame, from which the warping residuals are computed. The resulting residuals guide deeper block partitioning predictions.

A PyTorch-like implementation of the partitioning-adaptive warping is shown in Fig.~\ref{fig:code_pwarp}. For simplicity, we present the calculation only for a single reference frame. Visualization results are provided in Fig.~\ref{fig:viz_pwarp}. Using the \textit{Marketplace} sequence as an example, subfigure (a) shows the unprocessed predicted QT depth map, which exhibits strong local consistency due to the iterative top-down and bottom-up processing. Subfigures (b) and (c) display the original and partitioning-adaptive optical flow, respectively. The latter closely resembles block-based motion vector fields, as it incorporates the predictions from the QT depth map. Finally, subfigures (d), (e), and (f) present the inter-frame residuals, the warping residuals based on (b), and the warping residuals based on (c), respectively. Notably, (f) lies between (d) and (e), as it suppresses residuals caused by simpler motions captured by the QT depth map. This enables subsequent networks, such as the MTT mask Net and MTT Net, to better distinguish regions that require further partitioning from those that are already sufficiently partitioned.

\section{Detailed Training Process}\label{sec:training}
In this section, we provide details on constructing the training dataset, including the sources of the training data and the configuration for encoding the training sequences. We then discuss the detailed training strategy.

\subsection{Construction of Training Dataset}

We present the sources of our training dataset in Table \ref{tab:training_sequences}, which include BVI-DVC \footnote{\url{https://fan-aaron-zhang.github.io/BVI-DVC/}} \cite{bvi_dvc}, TVD \footnote{\url{https://multimedia.tencent.com/resources/tvd/}} \cite{TVD}, and UVG \footnote{\url{https://github.com/ultravideo/UVG-4K-Dataset/}} \cite{UVG}. For BVI-DVC, each sequence consists of 64 frames\footnote{In the latest NNVC reference software \url{https://vcgit.hhi.fraunhofer.de/jvet-ahg-nnvc/VVCSoftware_VTM/-/tree/VTM-11.0_nnvc/}, the BVI-DVC dataset has been expanded to 65 frames for dataset generation.}, divided into two Groups of Pictures (GOPs): frames $0\sim31$ and $32\sim63$. For TVD, each sequence contains 65 frames, with GOPs formed from frames $0\sim31$ and $33\sim64$. For UVG, frames are sampled at a 50-frame interval, generating multiple GOPs. In total, we obtain 668 GOPs, which are cropped to a resolution of $3840\times2160$ and converted to 8-bit depth. These 668 GOPs, originally at 4K resolution, are downsampled using bilinear interpolation via FFmpeg to generate GOPs at three additional resolutions: $1920\times1080$, $960\times544$, $480\times272$. We then use VTM 10.0 to encode these sequences with the configuration \textit{encoder\_randomaccess\_vtm\_gop32.cfg} to generate bitstreams. Herein, we set the intraperiod to 32.  During this process, we disable several fast algorithms in the VTM to obtain accurate block partitioning results, as detailed in Table \ref{tab:fast_tools}. Finally, the bitstreams are decoded, and the partition results can be exported and transformed into partition maps.
We use HDF5  to efficiently manage the training dataset, which includes the luma components of all training sequences, partition labels in the partition map format, and metadata for each frame, e.g. the bidirectional reference index of the current frame and the Slice QP of both the current and reference frames. The training dataset is available at \url{https://pan.baidu.com/s/1ZMPZqOcQS_gri_pzSq2vGA?pwd=tmxn}. Moreover, to further promote the following research, we also provide the code for generating the training dataset at \url{https://github.com/ustc-ivclab/IPM}, which includes: (1) the source code of modified VTM, which can output block partition statistics, and (2) a Python script for parsing the statistics into the partition map.

\begin{table}
    \centering
    \caption{4K Video Sequences for Constructing Training Dataset.}
    \label{tab:training_sequences}
    \resizebox{0.5\textwidth}{!}{\Large

        \begin{threeparttable}
    
        \begin{tabular}{c|cccc|c}
            \toprule
            \multicolumn{1}{c|}{Source}                               & \multicolumn{1}{c}{Resolution}                        & \multicolumn{1}{c}{\begin{tabular}[c]{@{}c@{}}\#seqs\end{tabular}} & \multicolumn{1}{c}{\begin{tabular}[c]{@{}c@{}}\#frames\end{tabular}} & \multicolumn{1}{c|}{\begin{tabular}[c|]{@{}c@{}}BitDepth\end{tabular}} & \begin{tabular}[|c]{@{}l@{}}\#GOPs\end{tabular} \\ 
            \midrule
            {BVI-DVC\cite{bvi_dvc}}                      & \cellcolor{gray!10}{3840$\times$2176}                  & \cellcolor{gray!10}{200}                                                          & \cellcolor{gray!10}{64}                                                        & \cellcolor{gray!10}{10}                                              & \cellcolor{gray!10}{400}                                                  \\ 
            TVD\cite{TVD}                                               & \multirow{1}{*}{3840$\times$2160}                 & \multicolumn{1}{c}{86}                                                           & \multicolumn{1}{c}{65}                                                        &                      {10}                           & {172}                                                  \\ 
            \multirow{2}{*}{UVG\cite{UVG}}                             &          \cellcolor{gray!10}{3840$\times$2160}             & \cellcolor{gray!10}{2}                                                            & \cellcolor{gray!10}{300}                                                       &                 \cellcolor{gray!10}{10}               & \cellcolor{gray!10}{12}                                                    \\ 
                                                                       &                  {3840$\times$2160}                                      & {14}                                                           & {600}                                                       &     {10}                                                           & {84}                                                   \\ 
                                                                       \midrule
            \multicolumn{5}{c|}{Total}                                                                                                                                                                                                                                                                                                             & \multicolumn{1}{c}{668}                                                  \\ 
            \bottomrule
        \end{tabular}%
    \end{threeparttable}  
    }
\end{table}

\begin{table}
    \centering
    \caption{The disabled fast tools in VTM during the training dataset preparation phase. The symbols \faCheck \ and \faTimes \ indicate the opening and closing of certain fast tools, respectively.}
    \label{tab:fast_tools}
    \resizebox{0.5\textwidth}{!}{

        \begin{threeparttable}
    
        \begin{tabular}{l|cc}
            \toprule
Fast Tools             & Default & Dataset Preparation \\ \midrule 
PBIntraFast                       & \faCheck       &   \faTimes                  \\ \rowcolor{gray!10}
ISPFast                            & \faTimes        &   \faTimes                  \\
FastMrg                            & \faCheck       &    \faTimes                 \\\rowcolor{gray!10}
AMaxBT                       & \faCheck       &    \faTimes                 \\
FastMIP                            &  \faTimes       &   \faTimes                  \\\rowcolor{gray!10}
FastLFNST                         &  \faTimes       &   \faTimes                  \\
FastLocalDualTreeMode             & \faCheck       &    \faTimes                 \\\rowcolor{gray!10}
ChromaTS                          & \faCheck       &    \faTimes                
 \\ \bottomrule
        \end{tabular}%
    \end{threeparttable}  
    }
\end{table}

\begin{table}[t]
    \centering
    \caption{Training Strategy. \includegraphics[width=0.35cm]{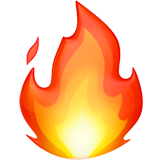} indicates trainable model parameters, while \includegraphics[width=0.35cm]{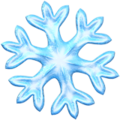} represents frozen model parameters. The three levels correspond to the three layers of the spatial pyramid, each associated with QT depth prediction, MTT mask prediction, and MTT depth/direction map prediction.
    }
    \label{tab:training_stage}
    \resizebox{0.5\textwidth}{!}{
        \begin{threeparttable}
            \begin{tabular}{ccc|cccc}
                \toprule
                \multicolumn{3}{c|}{Level} & \multirow{2}{*}{lr} & \multirow{2}{*}{Decay} & \multirow{2}{*}{BatchSize} & \multirow{2}{*}{Epoches} \\ 
                \cmidrule{1-3}
                S        & M        & L   &                     &                        &    &                     \\ \midrule
                \includegraphics[width=0.3cm]{Fig/fire.png}     &          &     &   $1\times10^{-3}$  & 0.98      &  160  &  500  \\\rowcolor{gray!10}
                \includegraphics[width=0.3cm]{Fig/snow.png}   & \includegraphics[width=0.3cm]{Fig/fire.png}  &   & $1\times10^{-3}$  &   0.98  &   32 &   300      \\
                \includegraphics[width=0.3cm]{Fig/snow.png}   & \includegraphics[width=0.3cm]{Fig/snow.png}   & \includegraphics[width=0.3cm]{Fig/fire.png}   &   $1\times10^{-3}$ &  0.98     &   8  &  300 \\\rowcolor{gray!10}
                \includegraphics[width=0.3cm]{Fig/fire.png}       & \includegraphics[width=0.3cm]{Fig/fire.png}      & \includegraphics[width=0.3cm]{Fig/fire.png}  &    $1\times10^{-4}$&    0.9   &   8   &  100 \\
                \bottomrule
            \end{tabular}
        \end{threeparttable}  
    }
    \begin{tablenotes}
    \item (1) "lr" refers to the learning rate. 
    \item (2) "Decay" refers to the decay rate of LR every 10 epochs.
    \item (3) "BatchSize" refers to the batch size for 4k sequences, while the batch sizes for $1920\times1080$, $960\times544$, and $480\times272$ are 4, 16, and 64 times those of 4k sequences, respectively. 
    \end{tablenotes}
\end{table}

\begin{figure*}
\label{fig:result2}
\centering 
    \includegraphics[trim=0cm 0cm 0cm 0cm, clip, width=0.9\textwidth]{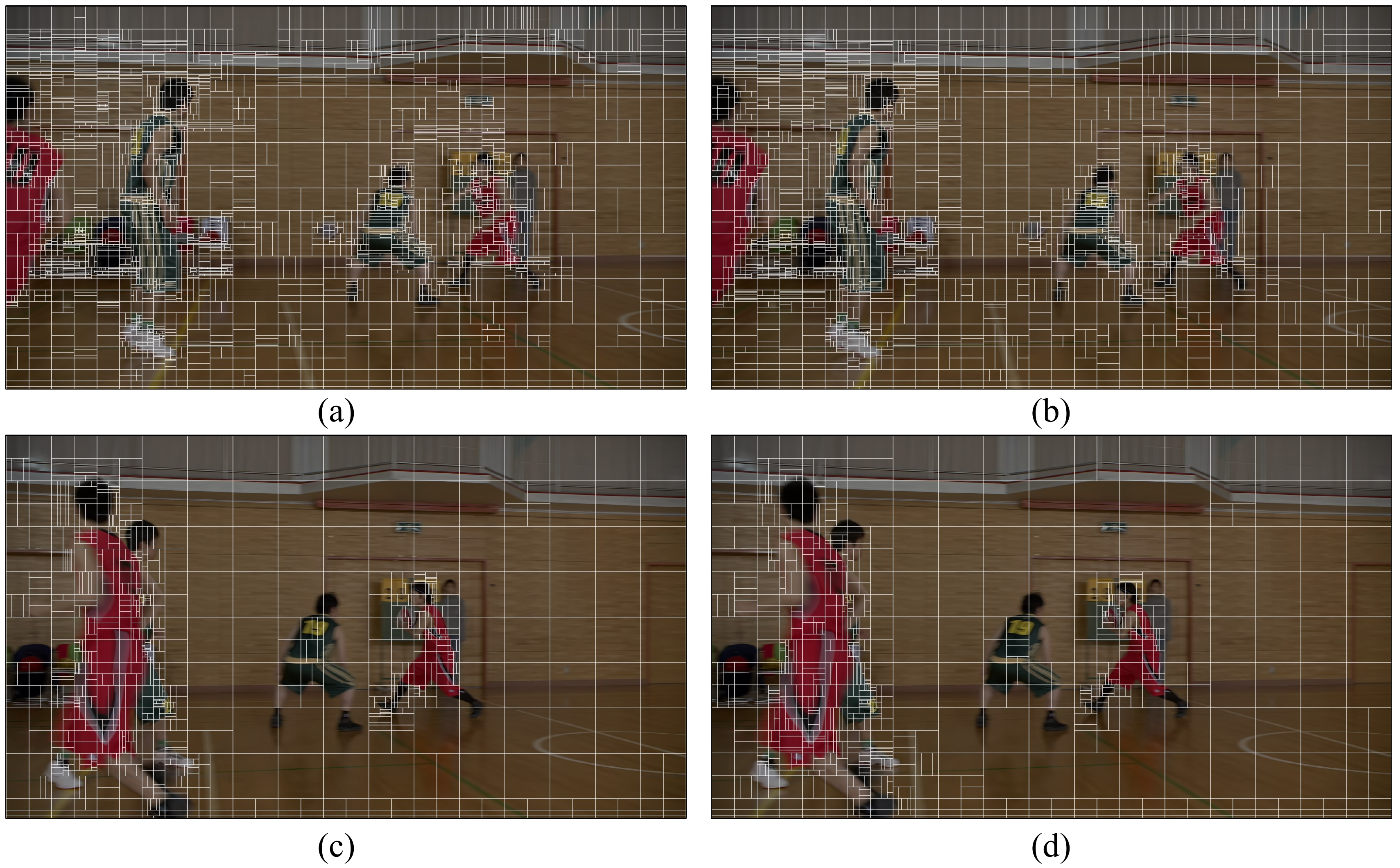}
	\caption{\textbf{Visualization of block partitioning results.} (a) and (b) correspond to the 15th frame of the \textit{BasketballDrive} sequence with a slice QP of 34. (a) represents the non-accelerated result, while (b) represents the accelerated result. (c) and (d) correspond tothe 3rd frame of the \textit{BasketballDrive} sequence with a slice QP of 41. (c) represents the non-accelerated result, while (d) represents the accelerated result. The results of our method are obtained at the trade-off point $L_0 (0.2,1)$.} 
	\label{fig:result1}
\end{figure*}
\begin{figure*} 
\centering
    \includegraphics[trim=0cm 0cm 0cm 0cm, width=0.9\textwidth]{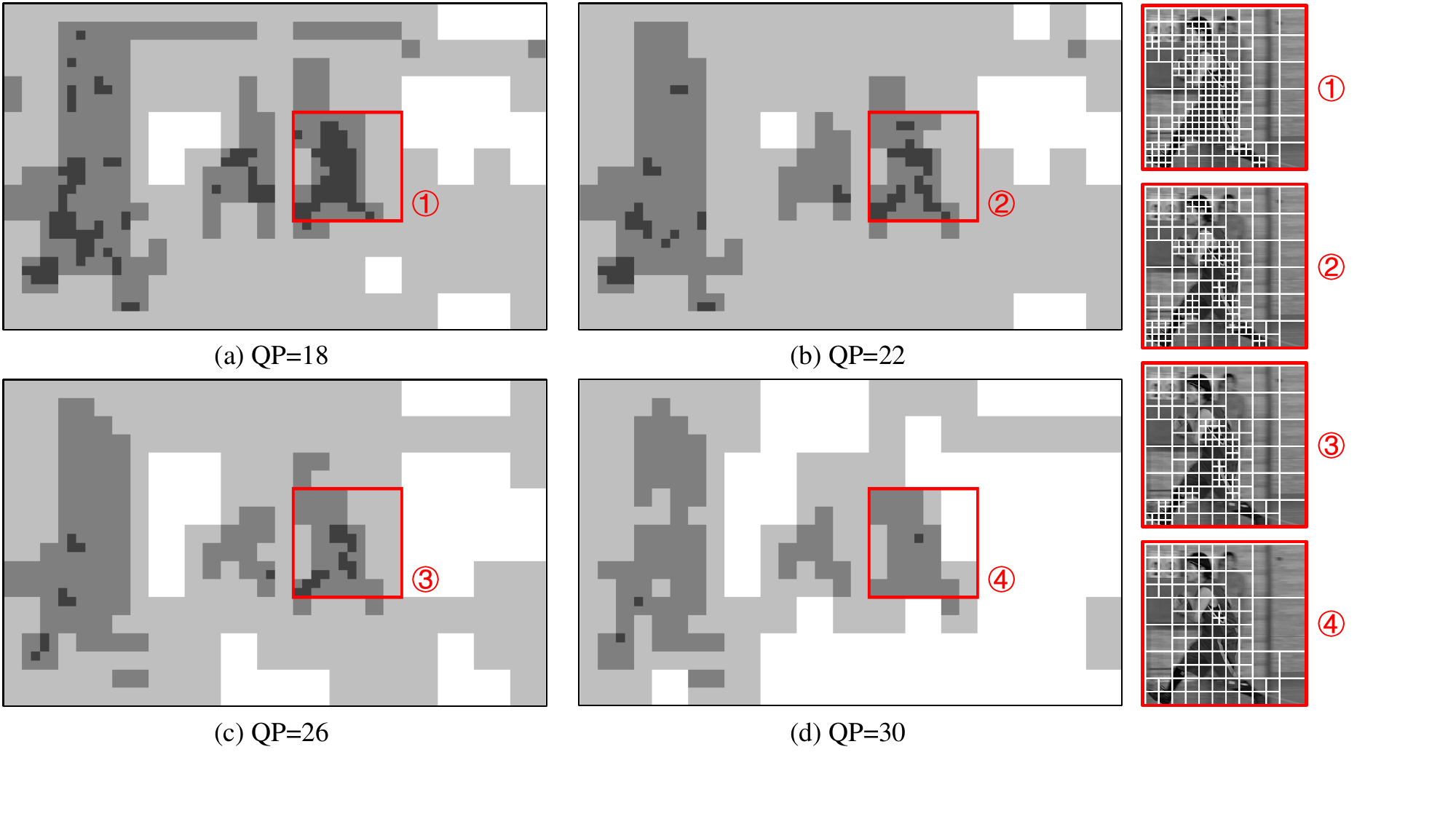}
    \caption{\textbf{Predicted QT depth map with varying input slice QP values.}
We select a specific CTU from the \textit{BasketballDrive} sequence as an example. Although the model is trained on a dataset with a basic QP of 22, the input slice QP can vary across a wide range. As a result, the predicted block partition structures also change, demonstrating that the QP modulation layer can effectively control the partitioning results.}
    \label{fig:varied_QP}
\end{figure*}

\begin{figure*} 
\centering
    \includegraphics[trim=0cm 0cm 0cm 0cm, width=0.9\textwidth]{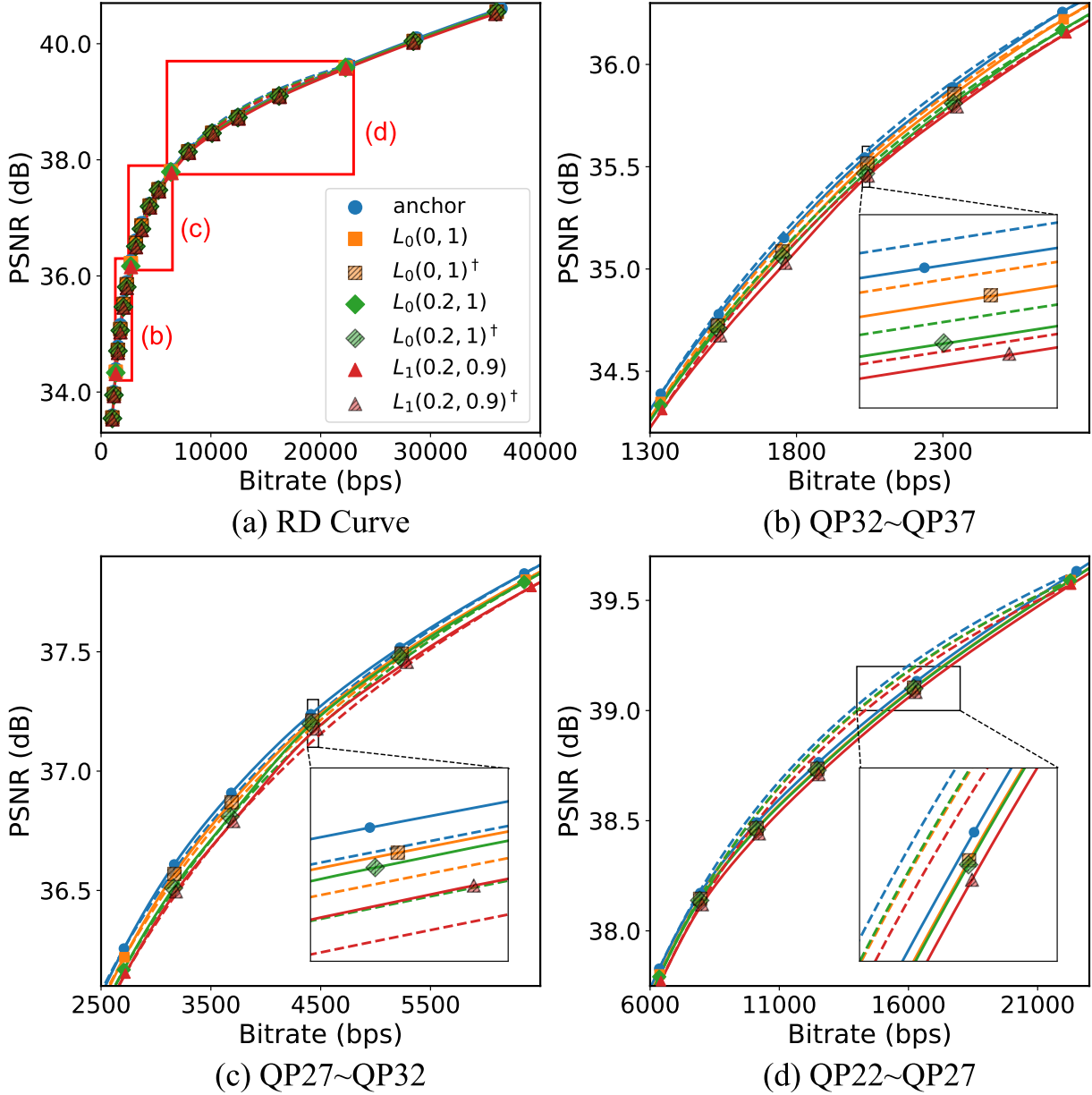}
    \caption{\textbf{Rate-Distortion Curve with Extended Base QPs.} Using the class B sequences as an example, we plot the RD points for base QPs ranging from 20 to 39 across three acceleration levels: $L_0(0,1)$, $L_0(0.1,2)$ and $L_1(0.2,0.9)$. The term ``anchor" refers to the original VTM encoder, while the superscript $^\dagger$ denotes the model applied to the extended QPs. 
    }
    \label{fig:refine_qp}
\end{figure*}

\subsection{Training Strategy}

Our training process is divided into two stages: separate training for each level of the spatial pyramid, and joint training for the overall scheme. In the first stage, we sequentially train the QT depth map prediction, MTT mask prediction, and MTT depth/direction map prediction, corresponding to three levels of the spatial pyramid.  We freeze the parts that have already been trained when training deeper levels, as detailed in Table \ref{tab:training_stage}.
Considering the different computational complexities at each level, we use different batch sizes at different stages and for different input resolutions to fully utilize computational resources. Detailed information on the batch sizes and epochs can be found in the table. Other configurations, such as the learning rate and its decay, remain unchanged across the training of each level. During the joint training stage, we set the initial learning rate to $10^{-4}$, with a decay factor of 0.9 every 10 epochs, and the batch size is set to 8.

\section{Detailed Evaluation Process}
\label{sec:evaluating}

In this section, we first introduce the strategy for encoding time measurement to obtain reliable encoding times. 
Next, we provide additional visual results. Finally, we present the detailed algorithm for the map tree-based post-processing.

\subsection{Encoding Time Measurement Setup}

In our experiments, we have observed that the measured encoding time may be unstable due to hardware/software fluctuations. Thus, we have used a proper setup to ensure the reliable results of encoding time. Details are following.
We use a statistical testing method based on the t-distribution assumption to obtain statistically reliable encoding times.
Specifically, we repeatedly test encoding times until they satisfy the following statistical test:
\begin{equation}
2 \cdot \frac{\sigma}{\sqrt{m}} \cdot t_{\alpha}(m-1) < \beta \cdot \bar{E}_{enc},
\end{equation}
where $m$ denotes the number of measurements, $\sigma$ represents the standard deviation of the measurement values, $\beta$ denotes the maximum deviation allowed for the encoding time, $\alpha$ represents the probability that the condition of $\beta$ is satisfied, $t_{\alpha}$ corresponds to the critical t-value of the Student's t-distribution, and $\bar{E}_{\text{enc}}$ denotes the average encoding time of the measurement series. We set $\beta$ to be 0.01 and $\alpha$ to 0.99, indicating a 99\% probability of a maximum deviation of 1\% for $\bar{E}_{enc}$ from the actual encoding time.
Furthermore, considering the high cost of each encoding time test, when the number of repeated encodings exceeds $M$ times, we employ Tukey's Fences 
to eliminate outliers based on quartiles and the interquartile range. Here, we empirically set $M$ to 8.

\subsection{Visualization}
We present visualizations of the block partitioning results produced by both the VTM encoder and the proposed method at the acceleration level \(L_0 (0.2, 1)\) in Fig.~\ref{fig:result1}. Specifically, we select the 15th and 3rd frames of the \textit{BasketballDrive} sequence, which have different Temporal IDs. Our method exhibits a block partitioning structure that closely resembles the ground-truth results produced by the VTM encoder.
Additionally, Fig.~\ref{fig:varied_QP} shows the predicted QT depth maps under varying slice QPs. Although the model is trained with a baseline QP of 22, it generalizes well to a wide range of input slice QPs. As the input QP varies, the granularity of the predicted block partition structures adapts accordingly, demonstrating that the QP modulation layer effectively controls the partitioning behavior.

\subsection{Map Tree-Based Post-processing Algorithm}
We present the detailed processing workflow of the map tree-based post-processing algorithm, as shown in Algorithm \ref{alg:post-process}. It inherits from the algorithm proposed in \cite{fal2023}, which transforms the prediction results of networks into a partition structure compliant with the VVC standard.

\floatname{algorithm}{Algorithm}  
\renewcommand{\algorithmicrequire}{\textbf{Input:}}  
\renewcommand{\algorithmicensure}{\textbf{Output:}}  

\begin{algorithm}	
	\caption{Generate the Optimal Map Tree\\
		\textbf{Description of intermediate variables:}\\
		$ QD_p $: Predicted QT depth map\\
		$ MD_p $: Predicted MTT depth maps\\
		$ MDir_p $: Predicted MTT direction maps }  
	\label{alg:post-process}
	\begin{algorithmic}[1]  
		\Require
		Initial $ MapNode $
		\Statex \ \quad Members are set to zero or empty set.
		\Ensure 
		Map tree
		\State \textbf{DataStructure} $ MapNode $:
		\State \ \ \ \ $ curQD $, $ curMD $, $ curMDir $,
		\State \ \ \ \ $ treeDepth$,
		\State \ \ \ \ $ children $, $ cus $  
		\Statex \ \ \ \ /* $ cus $ means all CUs of the current CTU */
		\Statex 
		\Function {GenerateMapTree}{$ node $}  
		\If{$ node.treeDpeth \ge maxTreeDepth $}
		\State \Return
		\EndIf 
		\State $ modeSet = \varnothing$
		\For{each $ cu $ in $ node.cus $}
		\State $ modeSet_{cu} = $ \Call{GetCandidateModes}{$ cu $}
		\Statex \qquad\quad/* Pruning the tree*/
		\State $ modeSet = modeSet \times modeSet_{cu} $
		\Statex \qquad\quad/* $ \times $ means Cartesian product */
		\EndFor
		\For{each $ mode_{cus} $ in $ modeSet $}
		\State Create child node based on $ mode_{cus} $: $ node_{c} $
		\State \Call{GenerateMapTree}{$ node_{c} $}
		\State $ node.children = node.children \cup \{node_{c}\} $
		\EndFor
		\EndFunction
		\Statex
		
		\Function {GetCandidateModes}{$ cu $}
		\State $ modeSet_{cu} = \varnothing$
		\State Generate a temporary QT depth map: $ QD_t $ 
		\State $ C = QD_t - QD_p $
		\If{$ number\ of\ negative\ values\ in\ C \le th_{qt} $}
		\State $ modeSet_{cu} = modeSet_{cu} \cup \{qt\} $ 
		\EndIf
		\For{each $ mode $ in $ \{no, bth, btv, tth, ttv\} $}
		\State According to $ mode $, generate a temporary  partition layer: $ MD_t $, $ MDir_t $ 
		\State $ error = ||MD_t-MD_p[curMTTdepth]||_1 +||MDir_t-MDir_p[curMTTdepth]||_1 $
		\If{$ error \le th_{mtt} $}
		\State $ modeSet_{cu} = modeSet_{cu} \cup \{mode\} $
		\EndIf
		\EndFor
		\State\Return{$ modeSet_{cu} $}
		\EndFunction

	\end{algorithmic}  
\end{algorithm}

\end{document}